\documentclass[alpha-refs]{wiley-article}

\usepackage{color}
\usepackage{ulem}

\usepackage{siunitx}
\usepackage{multirow}
\usepackage{graphicx}

\papertype{Original Article}
\paperfield{Journal Section}

\title{Spectral scaling of unstably-stratified atmospheric flows: understanding the low frequency spread of Kaimal model}

\abbrevs{MOST, Monin-Obukhov similarity theory; ML, Mixed Layer; TKE, Turbulence Kinetic Energy.}

\author[1\authfn{1}]{Claudine Charrondière}
\author[1\authfn{1}]{Ivana Stiperski}

\contrib[\authfn{1}]{Equally contributing authors.}

\affil[1]{Department of Atmospheric and Cryospheric Sciences, University of Innsbruck, Innsbruck, Austria}

\corraddress{Ivana Stperski,\\ Department of Atmospheric and
Cryospheric Sciences,\\ University of Innsbruck,
Innrain 52f, Innsbruck, Austria}
\corremail{ivana.stiperski@uibk.ac.at}

\fundinginfo{These results are part of a project that has received funding from the European Research Council (ERC) under the European Union’s Horizon 2020 research and innovation programme (Grant agreement No. 101001691).}

\runningauthor{Charrondière and Stiperski}

\begin{document}
\maketitle
\begin{abstract}

For more than five decades, a notable spread in the low frequencies of velocity and temperature spectra, scaled using inertial subrange properties and Monin-Obukhov similarity theory, has been observed in unstable stratification. A large ensemble of 14 datasets, over relatively simple terrain (from flat and homogeneous to gentle slope or valley floor) and very complex terrain (steep slope, crater rim, mountain top), are used to assess the reasons of this low-frequency behaviour.

Turbulence anisotropy is shown to be the primary factor influencing the spectral energy at the largest scales and the spectral peak position of streamwise and spanwise velocity spectra, with stability acting as a secondary factor. On the contrary, the low-frequency behaviour of surface-normal and temperature spectra is dominated by stability effects, while turbulence anisotropy plays a secondary role.
These observations are valid over both simple and complex terrain, even though variability of the largest scales of complex terrain datasets integrate other processes not included in turbulence anisotropy or stability conditions. 

Using a combination of non-dimensional turbulence parameters - temperature and velocity variances as well as dissipation of turbulence kinetic energy and of temperature variance - and of a semi-empirical model provided in the literature, we are able to describe the behaviour of the velocity and temperature spectra with only turbulence anisotropy and stability as input parameters.

Finally, the paper provides some insights into the nature of anisotropic eddies, as well as the relationship between the boundary layer height and the spectral behaviour.

\keywords{Atmospheric Boundary Layer, Complex Terrain, Monin-Obukhov Similarity Theory, Surface Layer, Turbulence Anisotropy}
\end{abstract}

\section{Introduction}

Spatial structure of atmospheric boundary layer is tightly coupled to the spectral characteristics of turbulence \citep{tennekes1972first}. Turbulence spectra, thus, provide invaluable information on the nature and structure of atmospheric turbulence, in particular on how energy (or variance) is distributed among the different scales of motions and how these scales interact with each others. In fact, spectral analysis is widely used in studying atmospheric boundary layer flows, especially in terms of understanding the process of turbulence cascade \citep[e.g.,][]{Kolmogorov1941a,Kolmogorov1941b,Stiperski2021}, 
the size and structure of most energetic eddies \citep[e.g.,][]{salesky2018buoyancy,Stiperski2020,stiperski2021convective,sun2020understanding, lan2022turbulence}, the departure of complex flows from canonical conditions \citep[e.g.,][]{Rotach1995,villani2003turbulence, Chamecki2004, babic2018}, the impact of large-scales atmospheric motions on the local flow \citep{dupont2022influence}, 
in testing the numerical schemes of weather prediction models and large eddy simulations \citep[e.g.,][]{sukoriansky2005application,wilczek2015spatio, gibbs2014comparison, dairay2017numerical}, as well as in providing a theoretical framework for widely used similarity scaling relations \citep[e.g.,][]{gioia2010spectral,katul2011mean, katul2013mean, katul2014two, banerjee2015revisiting,li2021keyps}. 

Surface-layer turbulence spectra are also one of the turbulence statistics that follow the Monin-Obukhov similarity theory \citep[MOST,][]{monin1954basic}. The first and still most widely used spectral scaling framework was proposed by \cite{kaimal1972spectral}, who suggested scaling relations for the spectra of velocity and temperature, as well as co-spectra of momentum and sensible heat fluxes. 
The spectral scaling was based on the Kolmogorov's hypothesis \citep{Kolmogorov1941b} such that the high-frequency part of the scaled spectra collapse in the inertial subrange, allowing a study of the low-frequency part of the spectra under MOST. 
Using data from the Kansas field experiment over horizontally homogeneous and flat terrain \citep{haugen1971experimental}, \cite{kaimal1972spectral} showed that under stable stratification the spectral energy at low frequencies is driven by MOST stability parameter $\zeta = z/L$, where $z$ is the height above the surface and $L$ the Obukhov length, defined as \citep{stull1988introduction}:
\begin{equation}
    L = -\frac{\overline{\theta}}{\kappa g} \frac{u_*^2}{T_*}.
\end{equation}
Here, $\overline{\theta}$ is the mean potential temperature, $g=9.81$~m~s\textsuperscript{-2} is the effective gravity, $\kappa=0.4$ is the von Kármán constant, $u_* =\left( \overline{u'w'}^2+\overline{v'w'}^2 \right) ^{1/4}$ is the friction velocity and $T_* = - \overline{w'\theta'}/u_*$ is the temperature scale, where $\overline{w'\theta'}$ is the kinematic sensible heat flux, while $\overline{u'w'}$ and $\overline{v'w'}$ are the momentum fluxes.

Still, in unstable conditions, \cite{kaimal1972spectral} observed a large spread of the low-frequency data that did not follow a particular scaling, concluding that the spectra "tend to cluster in a random fashion". They only noted a small separation of the surface-normal velocity spectra linked to $\zeta$ when approaching neutral conditions, and offered no hypothesis to explain such a spread.
Later, \cite{olesen1984modelling} used an analytical model based on the same approach to derive scaling laws for the velocity spectra both in the low-frequency and high-frequency range, but also failed in predicting the behaviour of low-frequency spectral energy for horizontal velocity spectra in unstable conditions. Given that the integral of the spectrum yields the variance, the non-scaling of the low-frequency part of the horizontal velocity spectra in unstable stratification can be linked to the failure of MOST in explaining the scaling of horizontal velocity variances, even over flat terrain \citep[e.g.,][]{WyngaardCote1974,kader1990mean}. 

Some refinements of MOST have been developed since.
A mixed layer scaling approach of \cite{panofsky1977characteristics} recognised that the horizontal velocity variances are driven by inactive large-convective eddies, and thus that the mixed layer (ML) depth $z_i$ and not the measurement height should be the relevant scaling height.
\cite{mcnaughton2007scaling} extended this approach to spectra and co-spectra. For temperature spectra and heat flux co-spectra, the scaling was then improved by a combination of $z_i$ and the surface friction layer height \citep{laubach2009scaling} and explains the position and spectral density of the energy peak. This means that the spectral shape and density are related to both local \citep{kaimal1972spectral} and bulk \citep{laubach2009scaling} information.
Based on the fact that the flow is influenced both by buoyancy and shear in the majority of stability conditions (except highly convective/stable), \cite{tong2015multipoint} developed the Multi-Point Monin-Obukhov theory, where the Obukhov length serves not only as the vertical, but also as the horizontal length scale. Through their approach, they could explain the commonly observed spectral slopes. 

Recently, \cite{stiperski2018dependence}, \cite{stiperski2019} and \cite{stiperski2023generalizing} showed that bulk turbulence statistics, such as velocity variances, temperature variance and turbulence kinetic energy (TKE) dissipation rate can be reconciled under MOST if turbulence anisotropy is included as an additional scaling parameter. \cite{mosso2023flux} extended the analysis to scaled velocity and temperature gradients and obtained similar conclusion and scaling relations.
The novel scaling relations drastically improve scaling and allow it to be extended to complex terrain, where the basic assumptions of MOST (horizontally homogeneous and flat terrain, constant vertical flux profile, no subsidence) are clearly violated. \cite{stiperski2023generalizing} suggested that, for horizontal velocity variances, the scaling based on anisotropy outperforms the ML scaling of \cite{panofsky1977characteristics}, which implies that turbulence anisotropy incorporates the majority of the information on the boundary conditions driving the flows (including the ML height).
Given the difficulties of obtaining ML height from conventional turbulence measurements and the lack of routine boundary layer height observations, the aforementioned scaling proposed in \cite{mcnaughton2007scaling} or in \cite{laubach2009scaling} are not easy to apply. A local parameter such as turbulence anisotropy is thus preferred for understanding spectral behaviour. 
At the same time, spectral analysis can help shed more light on the nature of turbulence anisotropy and how it is impacted by terrain complexity. 

\bigskip
    
Here we explore whether new results of \cite{stiperski2023generalizing} can be extended to two-point statistics such as spectral scaling of \cite{kaimal1972spectral}, in order to explain the spread of the low frequency range of velocity and temperature spectra in unstable stratification. 
More specifically, we investigate the influence of turbulence anisotropy on the spectral density and the location of the spectral peak for datasets with varying degree of complexity. 
We first focus on canonical conditions where the influence of terrain is expected to be small, and then proceed to complex mountainous conditions to show how much our method can account for the complexity of turbulence. Finally we explore turbulence anisotropy in light of the newly developed spectral scaling. 

Section~\ref{Section_theory} focuses on describing the theoretical framework on which the paper is based. Section~\ref{Section_data} provides an overview of the 14 datasets used, as well as a description of the data processing. Section~\ref{Section_results} presents the results with a special focus on the influence of stability conditions and turbulence anisotropy  (Section~\ref{Subsection_clustering_STAB} and \ref{Subsection_clustering}), separation between turbulence anisotropy and atmospheric stability effects (Section~\ref{Section_separation_parameters}), analytical modeling (Section~\ref{Subsection_modeling}), and the influence of mountainous terrain (Section~\ref{Section_complex_terrain}), followed by a discussion on the contribution of the mixed layer height to spectral scaling in Section \ref{Section_discussion} and conclusions in Section~\ref{Section_conclusion}. 

\section{Theoretical framework} \label{Section_theory}

The spectral scaling proposed by \cite{kaimal1972spectral} rests on two scaling frameworks: the inertial subrange scaling as proposed by \cite{Kolmogorov1941a} and \cite{Kolmogorov1941b}, and similarity functions of MOST \citep{monin1954basic}. Given the observed non-constancy of fluxes with height in non-canonical conditions \citep[e.g.,][]{nadeau2013similarity,sfyri2018}, we use the local scaling approach \citep{nieuwstadt1984some,stiperski2023generalizing} and scale the spectra at each measurement level with the respective fluxes at that level. Note that the classic MOST is an asymptotic case of local scaling in conditions when fluxes are constant with height. 

\begin{figure*}
\centering
	\includegraphics[width = \linewidth]{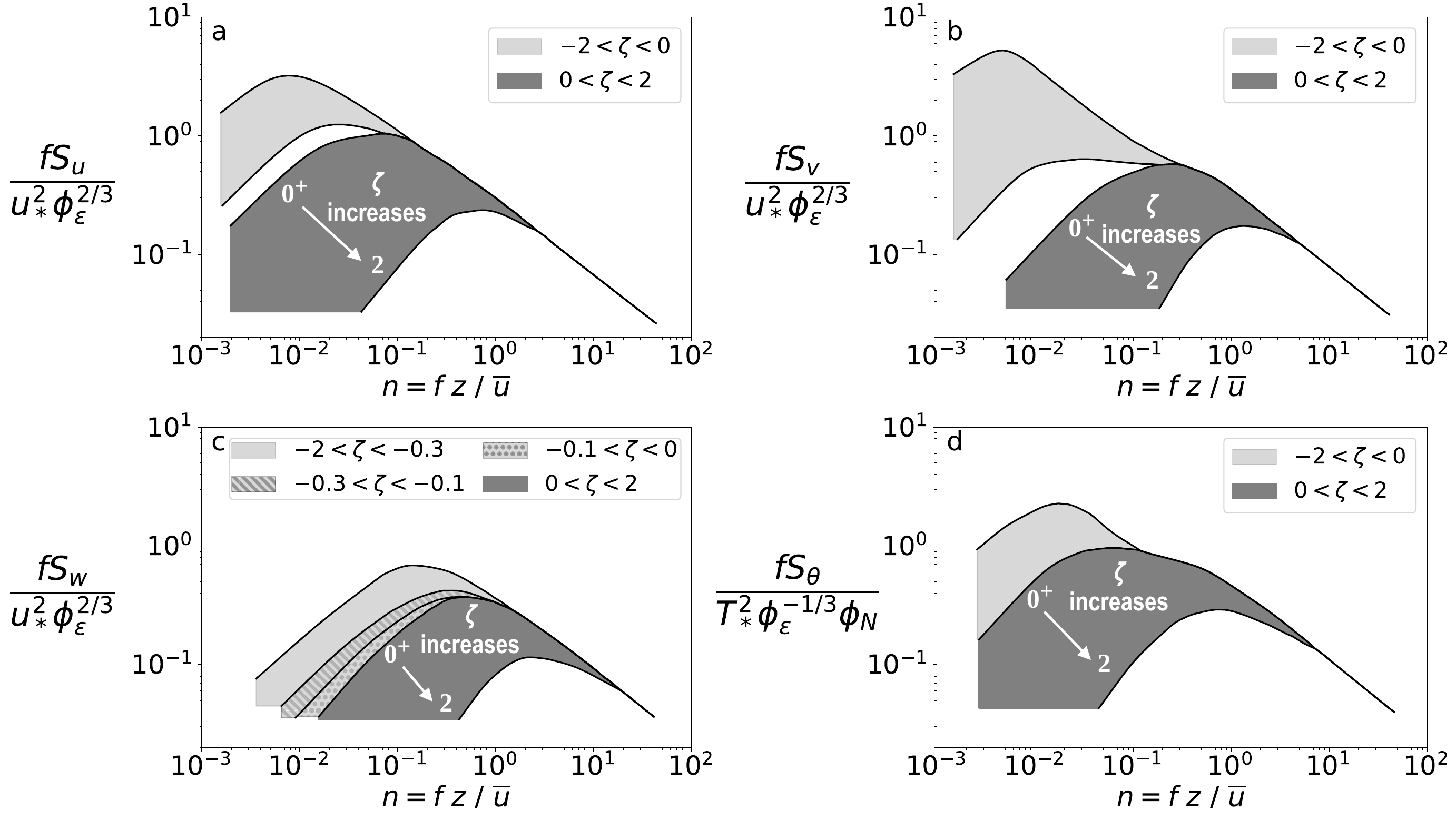}
	\caption{\label{Fig1} Scaled surface layer spectra for (a) streamwise, (b) spanwise, and (c) surface-normal velocity, and (d) temperature, as a function of stability parameter $\zeta$. Dark gray represents stable conditions, and light gray is for unstable conditions. Figure adapted from \cite{kaimal1994atmospheric}.}
\end{figure*}

According to Kolmogorov's hypotheses \citep{Kolmogorov1941a,Kolmogorov1941b}, the spectral density of the velocity spectra $S_{u_i}$ (with $u_1 = u$, $u_2 = v$ and $u_3 = w$ for streamwise, spanwise and surface-normal velocity components respectively) and temperature spectra $S_{\theta}$ in the inertial subrange are solely determined by the dissipation rates of TKE $\varepsilon$ and of half the temperature variance $N_{\theta}$ respectively, and can be written as \citep{pope2000turbulent, wyngaard2010turbulence}:
\begin{equation}
    S_{u_i} (k_x) = A_i \ \varepsilon^{2/3}\ k_x^{-5/3}, 
\end{equation}
\vspace{-1cm}
\begin{equation}
    S_{\theta}(k_x) = A_{\theta} \ N_{\theta}\  \varepsilon^{-1/3}\ k_x^{-5/3}.
\end{equation}
The wavenumber $k_x$ is parallel to the mean flow direction, $A_i$ is the Kolmogorov constant, equal to $A_1 \approx 0.55$ and $A_2 = A_3 = \frac{4}{3} A_1 $ for the three velocity components, and $A_{\theta} \approx 0.8$ is the Obukhov-Corrsin constant for the temperature variance. Following the Taylor's frozen turbulence hypothesis \citep{taylor1938spectrum}, the wavenumber spectrum can be converted from spatial to temporal domain, $k_x S(k_x) = f S(f)$, such that $k_x = 2\pi f / \overline{u}$, where $f$ is the frequency and $\overline{u}$ is the mean wind speed   \citep{wyngaard2010turbulence}. Finally, following MOST, the spectra can be written in their scaled form as \citep{kaimal1972spectral}:
\begin{equation}
\label{Eq_4}
     \frac{f S_{u_i}(f)}{u_*^2} = a_i \ \phi_{\varepsilon}^{2/3}\  n^{-2/3} \ ,
\end{equation}
\vspace{-.7cm}
\begin{equation}
\label{Eq_5}
    \frac{f S_{\theta} (f)}{T_*^2} = a_{\theta} \ \phi_N \ \phi_{\varepsilon}^{-1/3}\ n^{-2/3} \ .
\end{equation}
Here the spectra are scaled based on the key surface-layer parameters: $z$, $u_*$ and $T_*$. The scaled frequency $n=fz/\overline{u}$ and new values for the constants are $a_i = A_i(2\pi \kappa)^{-2/3}$, which gives $a_1 \approx 0.3, a_2 = a_3 \approx 0.4$, and similarly $a_{\theta} \approx 0.43$.  Finally, the MOST relations for TKE dissipation rate $\phi_{\varepsilon}$ and for half the temperature variance $\phi_N$ are defined as:
\begin{equation}
\label{Eq_6}
    \phi_{\varepsilon} = \dfrac{\kappa z \varepsilon }{u_*^3} \ ,
\end{equation}
\vspace{-.7cm}
\begin{equation}
   \phi_N = \dfrac{\kappa z N_{\theta}}{u_* T_*^2 } \ .
   \label{Eq_7}
\end{equation}
The scaled spectra based on Eqs.~\ref{Eq_4} and~\ref{Eq_5} collapse in the inertial subrange (Fig.~\ref{Fig1}). 

\cite{kaimal1972spectral} observed that the low frequency ranges of both velocity and temperature spectra vary according to the stability parameter for stable conditions ($\zeta>0$): the more stable the atmosphere, the less low frequency energy is present in the scaled spectra, and the spectral peak is found at higher $n$, such that energy containing eddies decrease in size and energy content as stratification increases. However, for $\zeta<0$, no such trend could be inferred. The only apparent separation between spectra according to $\zeta$ in unstable conditions could be discerned for surface-normal velocity in close to neutral stratification $-0.3 \leq \zeta \leq 0$ (Fig.~\ref{Fig1}c).

To derive a universal model for the spectra in stable conditions, \cite{kaimal1972spectral} studied the link between the scaled frequency of the spectral peak $n_m$ and $\zeta$. Using $n_m = f(\zeta)$, the spectra can then be described by a universal shape for stable atmospheric conditions. However, for horizontal velocity components in unstable atmospheric conditions, there is no clear trend of $n_m$ according to $\zeta$. Some studies took a constant $n_m$ \citep[e.g.,][]{kaimal1973turbulence,caughey1977boundary}, but the low-frequency spectral density of unstable horizontal velocity components was then not captured. 
This approach was employed later by \cite{olesen1984modelling}, who successfully reproduced spectra from the Kansas experiment in the stable conditions using $n_m$ proportional to the scaled velocity gradient $\phi_m$, taking the MOST relations for $\phi_m$ and $\phi_{\varepsilon}$. In unstable conditions, \cite{olesen1984modelling} assumed that $n_m= f(\zeta)$ only for small values of $\zeta$, and that it rapidly tends to a constant when stratification approaches the free convective limit. This reasoning is able to explain the behaviour shown in \cite{kaimal1972spectral} for surface-normal velocity spectra. 

According to \cite{olesen1984modelling}, the scaled velocity spectra assume the following expression:
\begin{equation}
\label{Eq_8}
    \frac{f S_{u_i}(f)}{u_*^2} = \frac{A_{u_i}
 n^{\gamma}}{(1+B_{u_i}n^{\alpha})^{\beta}},
\end{equation}
where parameters $A_{u_i}$ and $B_{u_i}$ determine the position of the spectra on the ordinate, while $\alpha$, $\beta$ and $\gamma$ influence its shape by providing asymptotic behaviour both at low and high frequencies.
Following Eq.~\ref{Eq_4}, the inertial subrange asymptote $(n \gg 1)$ of Olesen's model, as described in Eq.~\ref{Eq_8}, imposes two conditions on the free parameters:
\begin{equation}
\label{Eq_9}
    \gamma-\alpha \beta = -2/3 \ ,
\end{equation}
\vspace{-1cm}
\begin{equation}
\label{Eq_10}
    A_{u_i} = a_i \phi_{\varepsilon}^{2/3} B_{u_i}^\beta \ .
\end{equation}

Integrating Eq.~\ref{Eq_8} over the entire frequency range, \cite{olesen1984modelling} also derived an expression for the scaled velocity variance $\phi_{u_i}$, which, when combined with Eq.~\ref{Eq_10}, provides an analytical solution for $B_{u_i}$:
\begin{equation}
\label{Eq_11} 
B_{u_i} = \left(C_{u_i}\dfrac{\phi_{u_i}^2}{\phi_{\varepsilon}^{2/3}} \right)^{P} \textnormal{ with }  C_{u_i} = \dfrac{\alpha}{a_i} \dfrac{\Gamma(\beta)}{\Gamma\left(\dfrac{\gamma}{\alpha}\right)\Gamma \left(\dfrac{2}{3\alpha}\right)} \textnormal{ and } P = \dfrac{\alpha}{\alpha \beta-1} ,
\end{equation}
where $\Gamma$ is the gamma function, and the scaled velocity variance is defined as:
\begin{equation}
\label{Eq_12}
    \phi_{u_i}^2\ =\ \dfrac{\sigma_{u_i}^2}{u_*^2} \ .
\end{equation}
Equation ~\ref{Eq_11}, combined with Eq.~\ref{Eq_10}, allowed to compute the parameters $A_{u_i}$ and $B_{u_i}$, provided a set of values for $\alpha$, $\beta$ and $\gamma$. This approach has been used for example in \cite{tieleman1995universality}, and has the advantage of not requiring any information regarding the position of the maximum spectral energy $n_m$. \cite{tieleman1995universality} described the mathematical shapes of $A_{u_i}$ and $B_{u_i}$ for three sets of ($\alpha$, $\beta$, $\gamma$), and showed analytically that the velocity spectra are a function of $\phi_{u_i}$ only. 
Note that he used $\phi_{\varepsilon}= 1$, which is a good approximation only for near-neutral conditions. We thus do not expect this solution to perform well in convective conditions. 

Provided that $\phi_{\varepsilon} \neq 1 $ in the range of stability considered in the present study (down to $\zeta = -100$) we use the following adaptation of Eq.~\ref{Eq_8}:  
\begin{equation}
\label{Eq_13}
     \frac{f S_{u_i}(f)}{u_*^2 \phi_{\varepsilon}^{2/3}} = \frac{a_i B_{u_i}^{\beta} n^{\gamma}}{(1+B_{u_i}n^{\alpha})^{\beta}}
\end{equation}
One could adapt this semi-empirical model to temperature spectra consistently with Eq.~\ref{Eq_5}:
\begin{equation}
\label{Eq_14}
     \frac{f S_{\theta}(f)}{T_*^2 \phi_{\varepsilon}^{-1/3} \phi_N} = \frac{a_{\theta} B_{\theta}^{\beta} n^{\gamma}}{(1+B_{\theta}n^{\alpha})^{\beta}} ,
\end{equation}
\begin{equation}
    \textnormal { where }  \ \ B_{\theta}= \left(C_{\theta} \ \dfrac{\phi_{\theta}^2}{\phi_{\varepsilon}^{-1/3} \phi_N}\right)^{P}, 
\label{Eq_15}
\end{equation}
$C_{\theta}$ is defined as $C_{u_i}$ in Eq.~\ref{Eq_11}, replacing $a_i$ by $a_{\theta}$. The scaled temperature variance is defined as:
\begin{equation}
\label{Eq_16}
    \phi_{\theta}^2\ =\ \dfrac{\sigma_{\theta}^2}{T_*^2} \ .
\end{equation}

To include the information on turbulence anisotropy, we follow the work of \cite{stiperski2023generalizing}  who recently provided new empirical relations for $\phi_{u_i}$, $\phi_{\theta}$ and $\phi_{\varepsilon} = f(y_B, \zeta)$ that include the degree of anisotropy $y_B$ as a second dimensionless parameter (Table~\ref{table1}).
The degree of anisotropy is quantified by the anisotropy invariant analysis of the normalized anisotropy tensor $b_{ij}$ \citep{lumley1978computational}:
\begin{equation}
\label{Eq_17}
    b_{ij} = \dfrac{\overline{u_i'u_j'}}{2e} - \dfrac{1}{3}\delta_{ij},
\end{equation}
where $\overline{u_i'u_j'}$ is the Reynolds stress tensor, $e =0.5 \overline{u_l'^2}$ is the TKE, and $\delta_{ij}$ is the Kronecker delta. The degree of turbulence anisotropy is then quantified through the invariant $y_B$, determined from the smallest eigenvalue of $b_{ij}$, namely $\lambda_3$, following \cite{banerjee2007presentation}:
\begin{equation}
    y_B = (3\lambda_3+1) \frac{\sqrt{3}}{2}.
    \label{Eq_18}
\end{equation}
where $y_B=\sqrt{3}/2$ corresponds to pure isotropy, while $y_B=0$ is the most anisotropic state. For unstable stratification, the invariant $x_B$ was shown not to be important \citep{stiperski2019}. More details on the anisotropy analysis are provided in \cite{stiperski2018dependence}. The new scaling relations for the $\phi$ functions thus make it possible to include anisotropy into spectral scaling.

\begin{table}
\centering
\caption{Similarity scaling functions in the framework of MOST and according to the new framework of \cite{stiperski2023generalizing}}
\resizebox{\textwidth}{!}{%
\begin{tabular}{|c|l|l|}
\hline
\rowcolor[HTML]{C0C0C0} 
\textbf{} & \textbf{Monin-Obukhov similarity theory} & \textbf{Stiperski and Calaf (2023)} \\ \hline
$\phi_u$             & $2.55 \left[1-3\zeta\right]^{1/3}$ &  $[0.784-2.582 \log(y_B)] [1-3\zeta]^{1/3}$\\ \hline
$\phi_v$             &  $2.05 \left[1-3\zeta\right]^{1/3}$ & $[0.725-2.702 \log(y_B)] [1-3\zeta]^{1/3}$ \\ \hline
$\phi_w$             &  $1.25 \left[1-3\zeta\right]^{1/3}$ & $[1.119-0.019y_B-0.065y_B^2 + 0.028y_B^3] [1-3\zeta]^{1/3}$ \\ \hline
$\phi_{\theta}$      &  

\begin{tabular}[c]{@{}l@{}}$ 0.99\left[0.063-\zeta\right]^{-1/3}$, when $\zeta<-0.05$ \\ $0.015 \left|\zeta\right|^{-1} + 1.76 $, when $\zeta>-0.05$
\end{tabular} 

 & 
\begin{tabular}[c]{@{}l@{}}$ 1.07\left[0.05+|\zeta|\right]^{-1/3}$ \\ $+\left[-1.14+\left(0.017+0.217y_B\right)|\zeta|^{-9/10}\right]\left[1-\tanh{\left(10|\zeta|^{2/3}\right)}\right]$
\end{tabular} 
 \\ \hline
$\phi_{\varepsilon}$ & $\left[1-3\zeta\right]^{-1} - \zeta$ & \begin{tabular}[c]{@{}l@{}}$a\left[1-3\zeta\right]^{-1}-\zeta b$\\for $\varepsilon_u$: $a=0.371+1.811y_B$ \\\hspace{.78cm} $b=0.908+0.166y_B^{-1} + 0.009y_B^{-2}$ \\for $\varepsilon_w$: $a=0.219+1.846y_B$ \\\hspace{.82cm} $b=0.635+0.052y_B^{-1} + 0.006y_B^{-2}$ \end{tabular} 
\\ \hline
$\phi_N$             & $0.85\left[1-6.4\zeta\right]^{-2/3} / \phi_{\varepsilon}^{-1/3}$ &  \\ \hline
\end{tabular}%
}
\label{table1}
\end{table}

\section{Data}\label{Section_data}
\subsection{Datasets}

The 14 datasets used in the present study are summarized in Table~\ref{table2}. These datasets include: Advection Horizontal Array Turbulence Study \citep[AHATS,][]{horst2004hats}; Cabauw experimental site for Atmospheric Research \citep[Cesar,][]{verkaik2007wind}; Cooperative Atmosphere Surface Exchange Study 1999 \citep[CASES-99,][]{poulos2002cases}; five of the Innsbruck Box \citep[i-Box,][]{rotach2017investigating} turbulence towers (CS-VF0, CS-SF1, CS-NF10, CS-NF27, CS-MT21) together with the Im Hinteren Eis station; the NEAR and RIM towers from the Second Meteor Crater Experiment \citep[METCRAX II, stations MC-NEAR and MC-Rim,][]{lehner2016metcrax}; the F2 tower from the Mesoscale Alpine Programme MAP-Riviera experiment \citep[Riviera,][]{rotach2004turbulence} and the Central and West towers from Terrain-induced Rotor Experiment \citep[T-REX, stations T-RexC and T-RexW,][]{grubisic2008terrain}.

The stations are separated into two categories. The first category groups datasets that are closest to flat terrain (maximum slope is $3.25$°) : AHATS, Cabauw, CASES-99, CS-VF0, CS-SF1, MC-Near, T-RexC and T-RexW. We will refer to them as simple terrain thereafter. Note that CS-VF0, T-RexC and T-RexW are on the valley floor, meaning that they are surrounded by mountains even though they are on a relatively flat terrain. 
The second group is composed of datasets acquired either on steep slopes (from $10$° to $40$°, CS-NF10, CS-NF27 and Riviera), or in a highly complex configuration such on mountain tops or ridges (CS-MT21, Im Hinteren Eis, MC-Rim). We will refer to them as complex mountainous terrain. 

\begin{table}
\centering
\caption{Information on turbulence datasets used in the study and site characteristics}
\resizebox{\textwidth}{!}{
\begin{tabular}{|c|c|c|c|c|c|}
\hline
\rowcolor[HTML]{C0C0C0} 
\textbf{Experiment} & \textbf{Station name} &
  \textbf{Surface type} &
  \textbf{\begin{tabular}[c]{@{}c@{}}Terrain complexity\\ Slope angle\end{tabular}} &
  \textbf{\begin{tabular}[c]{@{}c@{}}Time range\\ available\end{tabular}} &
  \textbf{\begin{tabular}[c]{@{}c@{}}Measurement\\ heights (m)\end{tabular}} \\ \hline
\begin{tabular}[c]{@{}c@{}}AHATS\\ California (USA)\end{tabular} &
  AHATS &
  Fallow land &
  Flat &
  Jun-Aug, 2008 &
  \begin{tabular}[c]{@{}c@{}}$1.55$, $3.3$, $4.24$,\\ $5.53$, $7.08$, $8.05$\end{tabular} \\ \hline
\begin{tabular}[c]{@{}c@{}}Cesar\\ Netherlands\end{tabular} &
  Cabauw &
  Grass land &
  Flat &
  Jul-Oct, 2007 &
  $60$, $100$ \\ \hline
\begin{tabular}[c]{@{}c@{}}CASES-99\\ Kansas (USA)\end{tabular} &
  CASES-99 &
  Grass land &
  Flat &
  Oct, 1999 &
  \begin{tabular}[c]{@{}c@{}}$5$, $10$, $20$, $30$,\\ $40$, $50$, $55$\end{tabular} \\ \hline
 &
  CS-VF0 &
  Mixed agricultural &
  $0.5$° &
  Jan-Dec, 2015 &
  $4$, $8.7$, $16.9$ \\ \cline{2-6} 
 &
  CS-SF1 &
  Alpine meadow &
  $1$° &
  Jan-Dec, 2015 &
  $6.6$ \\ \cline{2-6} 
 &
  CS-NF10 &
  Alpine meadow &
  $10$° &
  Jan-Dec, 2015 &
  $6.65$ \\ \cline{2-6} 
 &
  CS-NF27 &
  Alpine meadow &
  $27$° &
  Oct-May, 2018 &
  $1.3$, $6.06$ \\ \cline{2-6} 
\multirow{-5}*{\begin{tabular}[c]{@{}c@{}}I-Box\\ Tirol (Austria)\end{tabular}} &
  CS-MT21 &
  High alpine vegetation &
  Mountain top &
  Apr-Oct, 2015 &
  $4.67$ \\ \hline
\begin{tabular}[c]{@{}c@{}}Im Hinteren Eis\\ Tirol (Austria)\end{tabular} &
  Im Hinteren Eis &
  Raw ground / snow &
  Mountain top &
  Jun-Oct, 2020 &
  $3.3$ \\ \hline
 &
  MC-NEAR &
  Desert &
  $1$° &
  Oct, 2013 &
  \begin{tabular}[c]{@{}c@{}}$10$, $15$, $20$, $25$, $30$, \\ $30$, $40$, $45$, $50$\end{tabular} \\ \cline{2-6} 
\multirow{-3}*{\begin{tabular}[c]{@{}c@{}}METCRAX II\\ Arizona (USA)\end{tabular}} &
  MC-Rim &
  Desert &
  Crater rim &
  Oct, 2013 &
  \begin{tabular}[c]{@{}c@{}}$5$, $10$, $15$, $20$, $25$, \\ $30$, $35$, $40$\end{tabular} \\ \hline
\begin{tabular}[c]{@{}c@{}}Map-Riviera\\ Switzerland\end{tabular} &
  Riviera &
  High alpine vegetation &
  $40$° &
  Aug-Sept, 1999 &
  $3.1$, $11$ \\ \hline
 &
  T-RexC &
  Desert &
  $<0.5$° &
  Mar-Apr, 2006 &
  $5$, $10$, $15$, $20$, $25$, $30$ \\ \cline{2-6} 
\multirow{-2}{*}{\begin{tabular}[c]{@{}c@{}}T-REX\\ California (USA)\end{tabular}} &
  T-RexW &
  Desert &
  $3.25$° &
  Feb-Apr, 2006 &
  $5$, $10$, $15$, $20$, $25$, $30$ \\ \hline
\end{tabular}}
\label{table2}
\end{table}

\subsection{Data processing}\label{Section_processing}

Turbulence data post-processing follows the procedure described in detail in \cite{stiperski2019} and used in \cite{stiperski2023generalizing}. All datasets were double rotated \citep{aubinet2012eddy} into the streamline coordinate system where $u_1$ is the streamwise, $u_2$ the spanwise, and $u_3$ the surface-normal velocity component. This coordinate system is standard in complex terrain studies, for which surface-parallel (streamwise and spanwise) direction, will diverge strongly from the horizontal. Turbulence statistics were determined based on the 30-min Reynolds averaging, with prior detrending, given that the present study focuses only on unstable conditions \citep[e.g.,][]{babic2017spectral, stiperski2018dependence}.
Furthermore, the data were required to be unstable at all measurement levels, and be acquired during daytime to avoid including nighttime counter-gradient fluxes into the statistics.

The following quality control was additionally applied. We kept only stationary data according to the stationarity test of \cite{foken1996tools}, with a threshold of $30$\%. The stationarity test was applied to the mean wind speed, as well as the sensible heat flux $\overline{w'\theta'}$ (except in near-neutral conditions, $\zeta>-0.05$, where the non-stationarity could be solely due to the small magnitude of the flux) and the turbulent momentum flux $\overline{u'w'}$ (except in highly convective conditions, $\zeta<-1$, where this flux has a small magnitude). Furthermore, we excluded 30-min segments for which the turbulent sensible heat flux $\overline{w'\theta'}<0.01$~m K s\textsuperscript{-1} at the corresponding measurement height, following \cite{franceschi2009analysis} and \cite{ nadeau2013similarity}. Such a small flux can be a sign of the (morning/evening) transition regime where $\overline{w'\theta'}$ integrates contributions from both unstable and stable regimes at different scales .

Spectral scaling of \cite{kaimal1972spectral} requires the knowledge of the TKE dissipation rate $\varepsilon$ and the dissipation rate of half the variance $N_{\theta}$ (Eqs.~\ref{Eq_4} and \ref{Eq_5}). These dissipation rates were determined from the second order structure function for streamwise, spanwise and surface-normal velocity variances \citep{Chamecki2004} and for temperature variances \citep{kiely1996convective}, under the usual assumptions of the existence of the inertial subrange \citep{Kolmogorov1941a}:
\begin{equation}
    \varepsilon_{u_i} = \left[\dfrac{D_{ii}(r)}{C_2^{u_i}}\right]^{3/2} r^{-1},
    \label{Eq_19}
\end{equation}
\begin{equation}
    N_{\theta} = \left[\dfrac{D_{\theta \theta}(r)}{C_2^{\theta}}\right] \varepsilon_u^{1/3} r^{-2/3},
    \label{Eq_20}
\end{equation}
where $r$ is the streamwise separation distance, and $C_2^{u} = 1.97$, $C_2^{v} = C_2^{w} = \frac{4}{3}C_u$, and $C_2^{\theta} = 3.2$. To assure that the second-order structure functions were evaluated within the inertial subrange, only lags for which the slope of the structure functions fell within the range of $2/3 \pm 10\%$ were used. 

Since some datasets cover a longer period or contain more measurement levels than others (Table~\ref{table2}), subsampling with replacement was used to build an ensemble of $8$ simple-terrain datasets, to ensure that each dataset is equally represented in the analysis. Each dataset within the ensemble was subsampled to have $500$ averaging periods within each bin of $\zeta$ or $y_B$. Two ensembles were thus independently created for spectra binned according to the logarithmically spaced $\zeta$ and linearly spaced $y_B$. The complex terrain datasets were evaluated independently. In the following figures, we present the ensemble medians and the interquartile ranges obtained from the bins of the logarithmically spaced normalized frequency $n$. Each bin of logarithmically spaced $n$ was required to have data from at least three datasets, and more than ten 30-min segments. 

\section{Results}\label{Section_results}
\subsection{Effect of stability on the spectra}\label{Subsection_clustering_STAB}

\begin{figure*}[ht]
\centering
	\includegraphics[width = .9\linewidth]{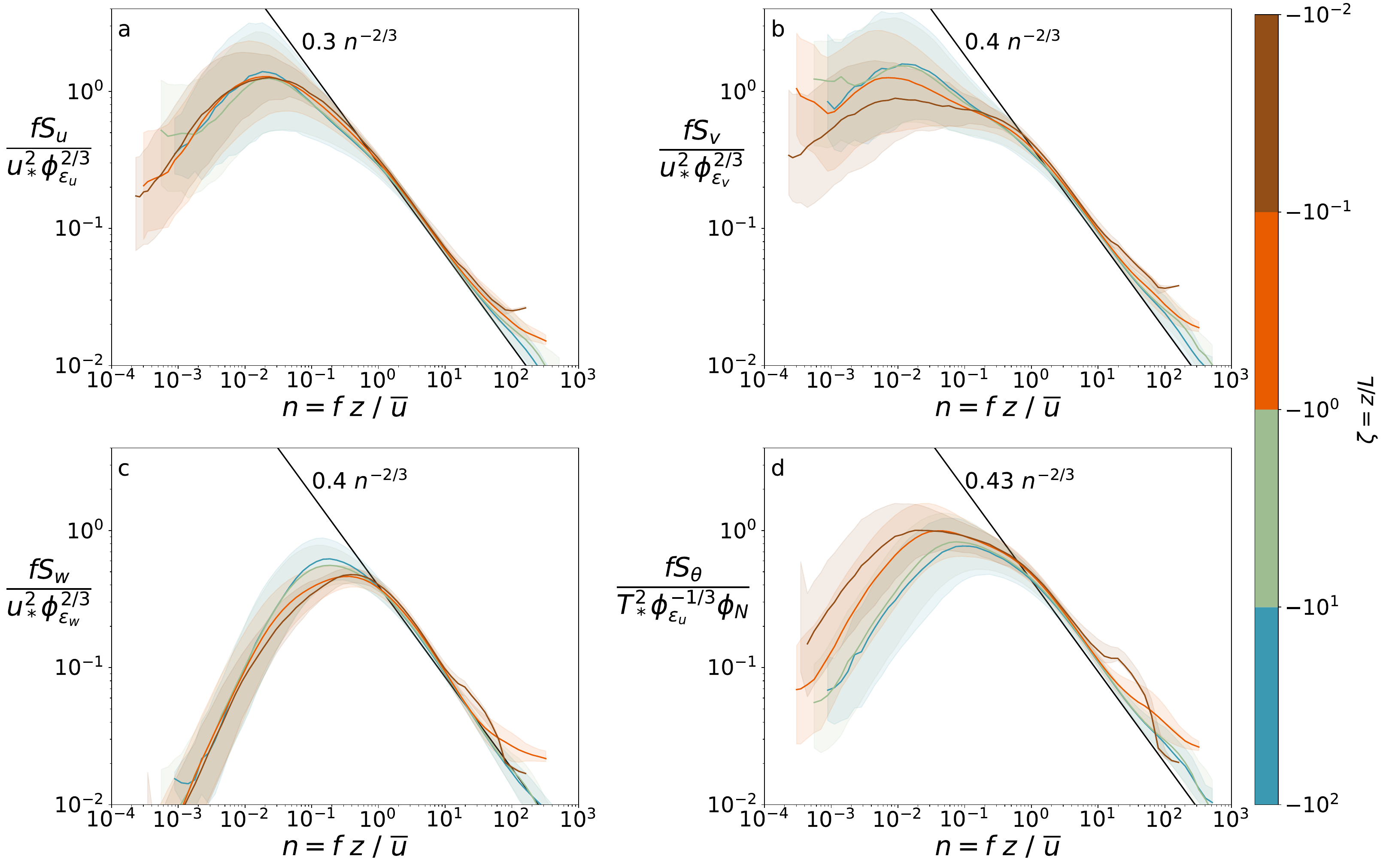}
	\caption{\label{Fig2} Ensemble of simple-terrain spectra of (a) streamwise, (b) spanwise and (c) surface-normal velocity and (d) temperature binned according to the logarithmically spaced stability parameter $\zeta$ (colour). Median of each $\zeta$ bin is shown with a solid colored line, and the interquartile range in shading. Solid black line corresponds to the spectral density expected in the inertial subrange according to Eqs.~\ref{Eq_4} and \ref{Eq_5}.}
\end{figure*}

We first explore the classical scaling of \cite{kaimal1972spectral} and bin the ensemble data according to $\zeta$ (Fig.~\ref{Fig2}). 
In agreement with \cite{kaimal1973turbulence}, we observe no dependence of the surface-parallel velocity spectra ($S_u$ and $S_v$) on $\zeta$, and only a small dependence of $S_w$. We, however, observe three major differences between our and \cite{kaimal1973turbulence}'s results. The low-frequency part of the spanwise spectra $S_v$ show reduced spectral density in the near-neutral conditions ($\zeta>-0.1$) and a plateau-like shape, compared to more unstable stratification. The existence of a plateau (observed also in individual datasets), indicates a continuum of eddy sizes contributing significantly to the total variance. In convective conditions the spectral peak of $S_v$ is better defined, although still broader than in the streamwise velocity spectra $S_u$. The spanwise velocity spectra $S_v$ also exhibits a slightly concave shape between the frequency of the energy peak and lowest frequency limit of the inertial subrange, especially visible for very anisotropic conditions, as already pointed out in \cite{kaimal1978horizontal}. 

The small dependence of $S_w$ on $\zeta$ stems solely from the dependence of the position of the spectral peak on $\zeta$. This peak is shifted from approximately $n=0.1$ in very convective conditions to approximately $n=0.7$ when $|\zeta| < 1$. This means that the surface-normal extent of the most energetic eddies is largest in convective conditions, as expected. However, we do not observe any impact of stability on the lowest frequencies.

The major difference between \cite{kaimal1973turbulence} results and ours arises for temperature spectra, which show a very clear dependence on $\zeta$. An increase of $|\zeta|$ is seen to correspond to a decrease of the low-frequency spectral energy and a shift of the spectral peak towards smaller scales. \cite{laubach2009scaling} also mentioned such a shift of the peak, with increasing instability during the day. 

\subsection{Effect of anisotropy on the largest scales of turbulence}\label{Subsection_clustering}

According to \cite{stiperski2023generalizing}, velocity and temperature variances are functions of both the stability parameter $\zeta$ and turbulence anisotropy $y_B$. Knowing that turbulence anisotropy is scale-dependent, and that the dominant anisotropy stems from the anisotropic forcing at energy-containing scales, the question arises whether $y_B$ can explain the lack of scaling of surface-parallel velocity spectra, as well as whether the low frequency contributions among the different degrees of anisotropy show consistent behaviour. 

\begin{figure*}[ht]
\centering
	\includegraphics[width = \linewidth]{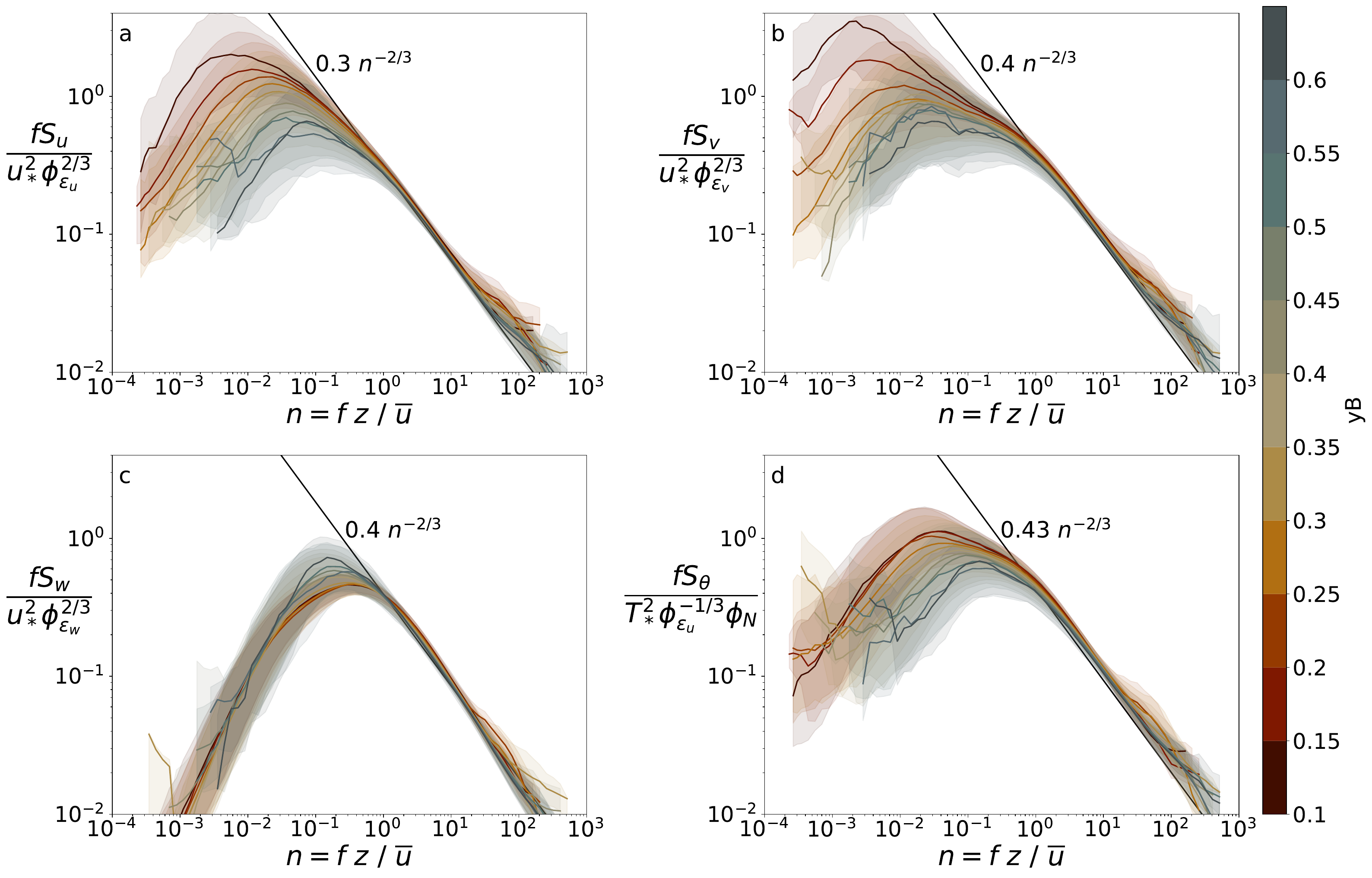}
	\caption{\label{Fig3}  The same as Fig. \ref{Fig2} but binned according to the degree of anisotropy $y_B$ (colour). }
\end{figure*}

To answer these questions, the velocity and temperature spectra are binned according to the degree of turbulence anisotropy, irrespective of $\zeta$ (Fig.~\ref{Fig3}). Spectra of surface-parallel velocities, $S_u$ and $S_v$, as well as temperature $S_{\theta}$, indicate a clear dependence on $y_B$ in the range of frequencies that do not scale under the \cite{kaimal1972spectral} scaling. This behaviour is consistent with the results of \cite{stiperski2023generalizing} for the velocity variances and shows that the degree of anisotropy is indeed the driving mechanism behind the non-scaling of surface-parallel velocity spectra. On the other hand, the surface-normal velocity spectra $S_w$ (Fig.~\ref{Fig3}c) show only a modest dependence on $y_B$ in terms of the energy at peak frequency: isotropic turbulence corresponds to a lower peak frequency which contains more energy than anisotropic turbulence. This dependence, however, is not uniform as an anisotropy threshold needs to be reached for $y_B$ to show effects on the spectral peak.

Figure~\ref{Fig3} also offers a deeper insight into the nature of turbulence eddies associated with the different degrees of anisotropy. The results show that quasi-isotropic turbulence (large $y_B$) is associated with smaller eddies that carry less energy in the surface-parallel, but more in the surface-normal direction than anisotropic turbulence, thus leading to an almost equal energy distribution in all the three directions. Furthermore, the peaks in the velocity spectra for isotropic turbulence occur at similar frequencies for all three velocity components, indicating that isotropic turbulence is also quasi-isotropic in eddy size. As the anisotropy increases ($y_B$ decreases), on the other hand, the horizontal size and the energy content of the dominant energy-containing eddies increases (the spectral peak is shifted towards lower frequencies) compared to more isotropic turbulence. In fact, the surface-normal velocity component shows strongest dependence on $y_B$ in quasi-isotropic range, while the surface-parallel velocity spectra show largest influence in anisotropic conditions. 

\subsection{Joint effects of turbulence anisotropy and stability}\label{Section_separation_parameters}

\begin{figure*}
\centering
\includegraphics[width = .7\linewidth]{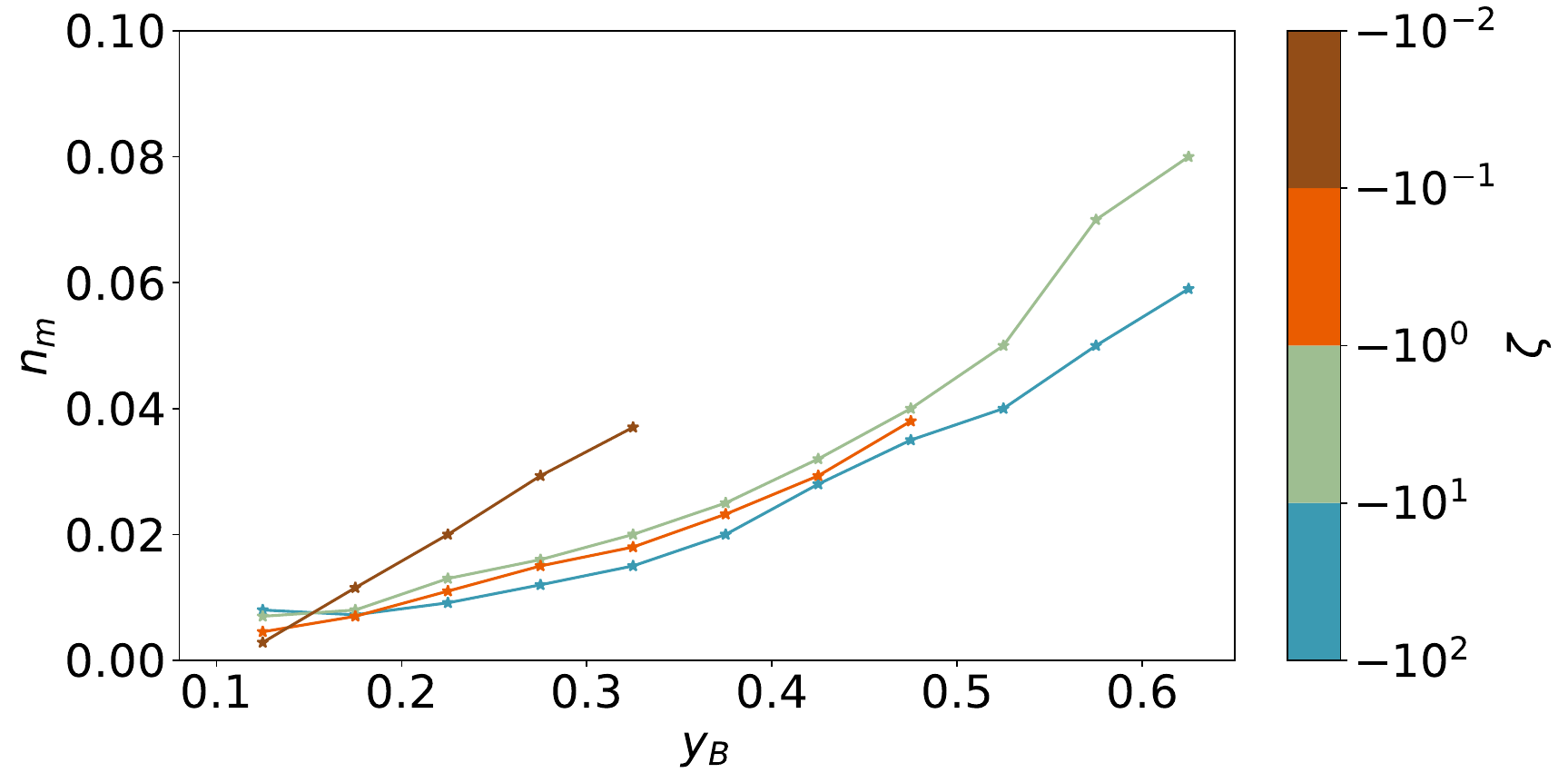}
\caption{\label{Fig10} Spectral peak normalized frequency as a function of turbulence anisotropy (abscissa) and stability (colored scale). Each point is computed from the median spectra of a given bin of $y_B$ and $\zeta$.}
\end{figure*}

\begin{figure*}
\centering
\includegraphics[width = \linewidth]{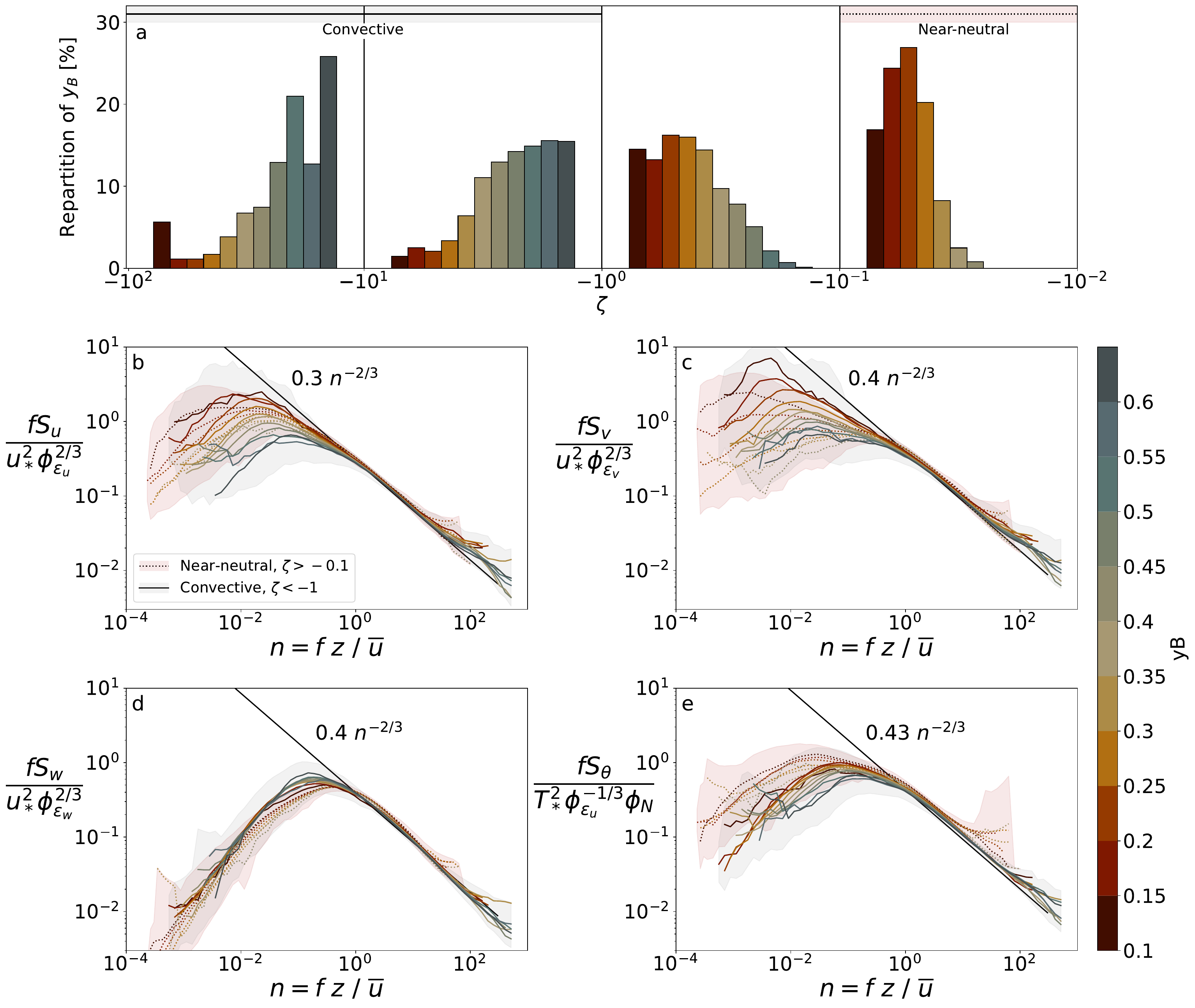}
\caption{\label{Fig4}  (a) Distribution of anisotropy state within each bin of $\zeta$ in the ensemble data. Spectra of (b) streamwise, (c) spanwise and (d) surface-normal velocity and of (e) temperature, binned with respect to turbulence anisotropy $y_B$ for convective (solid lines) and near-neutral (dotted lines) conditions. The shading includes the interquartile range for the whole stability conditions considered, regardless of turbulence anisotropy bins.}
\end{figure*}

Turbulence anisotropy, while defined only based on the Reynolds stress tensor (Eqs.~\ref{Eq_17} and \ref{Eq_18}), is the result of multiple physical processes and their interactions, including stability-related processes. 
To separate the effect of $y_B$ from the effect of $\zeta$ in the scaled spectra, we use the spectral peak frequency $n_m$ of $S_u$ as a marker. Figure~\ref{Fig10} shows $n_m$ as a function of $y_B$ for the four bins of $\zeta$ defined previously (cf. Fig.~\ref{Fig2}). The behaviour of $n_m$ differs significantly depending on whether conditions are near-neutral ($|\zeta|<0.1$) or more convective ($|\zeta|>0.1$). For near-neutral conditions, the spectral peak frequency increases linearly with $y_B$ while for convective conditions the growth is exponential. For very anisotropic conditions ($y_B<0.35$) thus the increase of $n_m$ with $y_B$ is weaker than in near-neutral conditions. In addition, for a given value of $y_B$, the spectral peak occurs at lower scaled frequencies $n_m$ than in near-neutral stratification. This feature is lost when examining ensemble spectra binned according to $\zeta$ (Fig.~\ref{Fig2}), since each stability bin contains different anisotropy types: very anisotropic turbulence develops preferentially in near-neutral conditions, while the most isotropic conditions are found only in convective conditions (Fig.~\ref{Fig4}a), as already observed in \cite{stiperski2018dependence} and \cite{stiperski2021convective} 

Figure~\ref{Fig4}b-e shows the velocity and temperature spectra binned according to turbulence anisotropy for two stability classes: near-neutral ($|\zeta|<0.1$) and highly convective ($|\zeta >1|$). 
The low-frequency separation of the streamwise and spanwise velocity spectra according to $y_B$ occurs in equal measure for near-neutral and convective conditions (Fig.~\ref{Fig4}b-c), with however several differences: (1) quasi-isotropic turbulence only occurs in convective conditions, while more anisotropic turbulence is present in all stability ranges as above-described; (2) for the most anisotropic data, the convective spectra show a well defined peak at higher frequencies than in near-neutral spectra, which instead show a plateau or a very broad peak with the maximum at lower frequencies; (3) spectral energy is higher but also more variable at low frequencies in convective conditions compared to near-neutral conditions.

The very small dependence of the surface-normal spectra on $y_B$ (Fig.~\ref{Fig3}c) is related to the associated stability conditions (Figs.~\ref{Fig2}c and \ref{Fig4}d). The most striking result is the small but clear increase of energy in the surface-normal direction with increasing $y_B$ in convective conditions. 
The temperature spectra (Fig.~\ref{Fig4}e), on the other hand, show only a limited variability according to $y_B$ for near-neutral stratification, but a clear separation according to anisotropy in convective conditions. The clear low frequencies binning of temperature spectra with $y_B$ in Fig.~\ref{Fig3}d is thus the signature of the repartition of anisotropic states within the stability range considered. 

Figure~\ref{Fig4} thus clearly highlights the different characteristics of anisotropic eddies under the two stability regimes. Anisotropic eddies under near-neutral stratification are predominantly driven by shear that injects energy into the streamwise (and spanwise) velocity component. The streamwise and spanwise spectra are thus characterised by larger surface-parallel size of the most energetic eddies that occur over a wider range of scales (Fig.~\ref{Fig4}b, c), and whose size scales with height ($n = 1$) in the surface-normal direction (Fig.~\ref{Fig4}d). 
The anisotropic eddies that occur under convective conditions, on the other hand, have surface-parallel spectra with a better defined peak that corresponds to smaller eddies in the surface-parallel, but larger in the surface-normal direction than in near-neutral conditions. The energy content of the most energetic eddies, however, is larger than under near-neutral stratification in all three directions, especially spanwise. 

\subsection{Analytical modeling of spectra}\label{Subsection_modeling}

Following the semi-empirical model of \cite{olesen1984modelling} described in Sect.~\ref{Section_theory}, the present Section intends to describe the spectral shape, with a function requiring only the knowledge of local parameters (namely $y_B$ and $\zeta$). To do so, we use Eqs.~\ref{Eq_13} and \ref{Eq_14} and take $\gamma = 1$, as widely used in micrometeorology \citep[cf. ][]{kaimal1972spectral, olesen1984modelling,tieleman1995universality}. 
As for $\alpha$ and $\beta$, there is no consensus on their value in the literature. In stable conditions, \cite{olesen1984modelling} used $\beta=5/3$ for all three velocity components. In near-neutral conditions, \cite{tieleman1995universality} differentiates between ideal terrain (flat, smooth and uniform), where $\beta = 1$ is the best fit, and complex terrain (heterogeneous surface roughness, topography conditions), where $\beta = 5/3$ is more suitable. 
According to \cite{kaimal1978horizontal}, a value of $5/3$ is also the best fit for the surface-normal velocity spectra in unstable conditions. 
In the present paper, we adapted $\beta$, without any a priori assumptions, to the value that fits best to the spectra.
We observe that $\beta=5/3$ is a good fitting parameter only for $S_u$, while $S_v$ and $S_w$ require respectively $7/3$ and $4/3$, and $S_{\theta}$ is best described by $\beta=2$.  The remaining exponent $\alpha$ is derived from Eq.~\ref{Eq_9}, knowing $\beta$ and $\gamma$. These coefficients, and the resulting parameters, used to derive the solution are summarized in Table~\ref{table3}. 

\begin{table}
\centering
\caption{List of coefficients used in the spectral model, Eqs~\ref{Eq_13} and \ref{Eq_14}}
\begin{tabular}{|ll|ll|ll|ll|}
\hline
\rowcolor[HTML]{EFEFEF} 
\multicolumn{2}{|c|}{\cellcolor[HTML]{EFEFEF}\begin{tabular}[c]{@{}c@{}}Streamwise velocity\\spectra $S_u$\end{tabular}}   
& \multicolumn{2}{c|}{\cellcolor[HTML]{EFEFEF}\begin{tabular}[c]{@{}c@{}}Spanwise velocity\\spectra $S_v$\end{tabular}} & \multicolumn{2}{c|}{\cellcolor[HTML]{EFEFEF}\begin{tabular}[c]{@{}c@{}}Surface-normal\\velocity spectra $S_w$\end{tabular}} & 
\multicolumn{2}{c|}{\cellcolor[HTML]{EFEFEF} \begin{tabular}[c]{@{}c@{}}Temperature\\spectra $S_{\theta}$\end{tabular}} \\ \hline
\rowcolor[HTML]{FFFFFF} 
\multicolumn{1}{|l|}{\cellcolor[HTML]{FFFFFF}
\begin{tabular}[c]{@{}l@{}}$\alpha = 1$\\ $\beta = 5/3$\\ $\gamma = 1$\end{tabular}} 
& 
\begin{tabular}[c]{@{}l@{}}$C_u = 2.24$ \\ $P= 1.5$\end{tabular}  
& 
\multicolumn{1}{l|}{\cellcolor[HTML]{FFFFFF}  
\begin{tabular}[c]{@{}l@{}}$\alpha = 5/7$\\ $\beta = 7/3$\\ $\gamma = 1$\end{tabular}}      
&     
\begin{tabular}[c]{@{}l@{}}$C_v = 3.09$ \\ $P= 1.07$\end{tabular}     & 
\multicolumn{1}{l|}{\cellcolor[HTML]{FFFFFF}      
\begin{tabular}[c]{@{}l@{}}$\alpha=5/4$\\ $\beta=4/3$ \\ $\gamma=1$\end{tabular}} 
&              
\begin{tabular}[c]{@{}l@{}}$C_w = 1.94$ \\ $P= 1.875$\end{tabular}
& \multicolumn{1}{l|}{\cellcolor[HTML]{FFFFFF}                 
\begin{tabular}[c]{@{}l@{}}$\alpha=5/6$\\ $\beta=2$ \\ $\gamma=1$\end{tabular}}
&  \begin{tabular}[c]{@{}l@{}}$C_{\theta} = 1.8$ \\ $P= 1.25$\end{tabular}                \\ \hline
\end{tabular}
\label{table3}
\end{table}

\begin{figure*}
\centering
	\includegraphics[width = .88\linewidth]{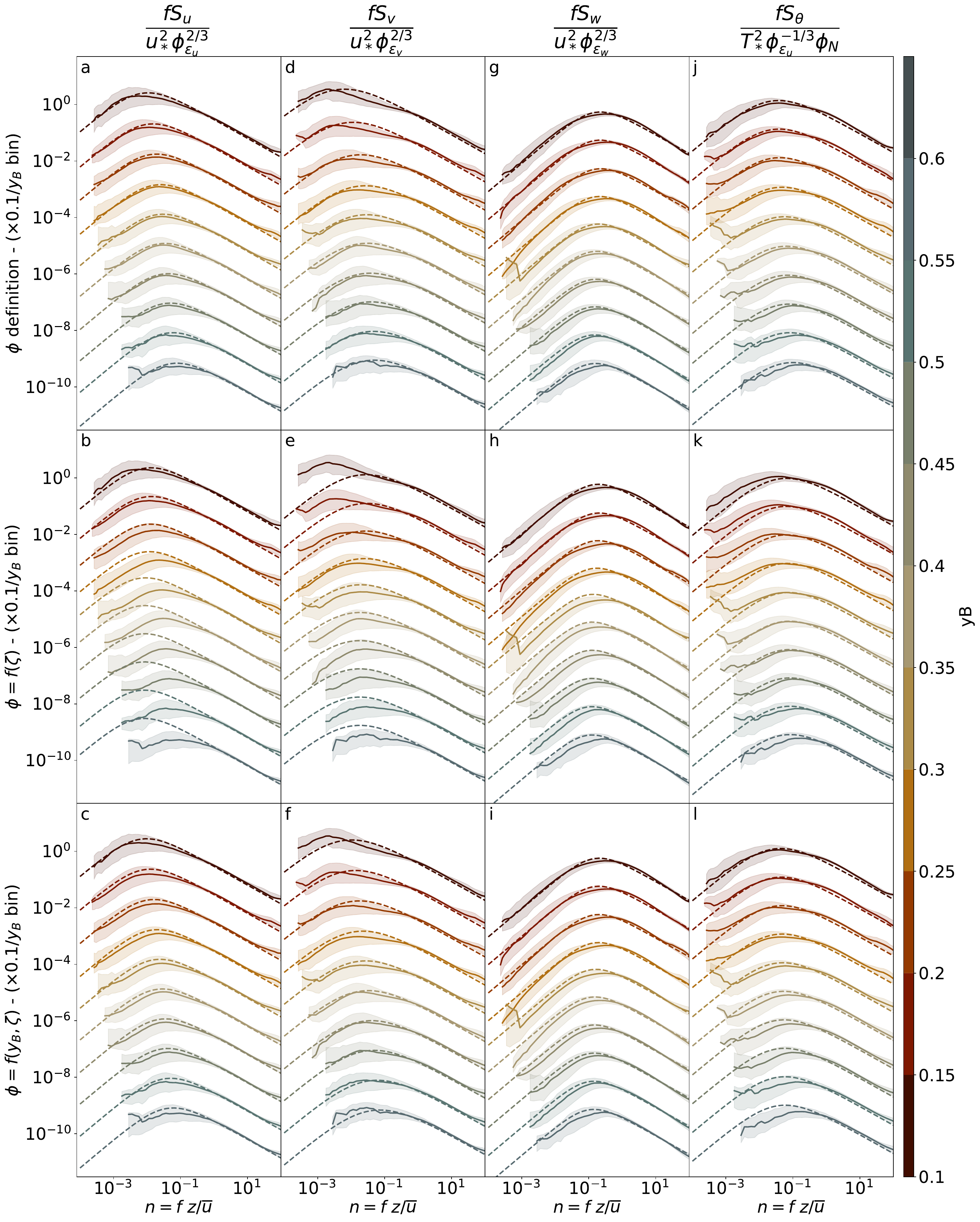}
	\caption{\label{Fig5} Scaled (a-c) streamwise, (d-f) spanwise, (g-i) surface-normal velocity spectra and (j-l) temperature spectra. In each subplot, the solid lines represent the median of a range of turbulence anisotropy from the ensemble data of simple terrain dataset. The dashed lines are the solution of Olesen's model, computed for each individual spectrum before taking their median. The shading is the interquartile range for each anisotropy bin (scale on the right for the corresponding values of $y_B$) In the first row, the solution is computed using the definition of $\phi$ (Eqs.~\ref{Eq_6}, \ref{Eq_7}, \ref{Eq_12} and \ref{Eq_16}), the second row when using the $\phi$ functions according to MOST, and third row when using $\phi$ functions according to \cite{stiperski2023generalizing} (Table~\ref{table1}). To facilitate the reading, spectra in each subplot are shifted down compared to each other by one decade, where the upper-most spectrum is in the correct position.}
\end{figure*}

Figure~\ref{Fig5} shows the velocity and temperature spectra along with the model predictions. The model, given by Eq.~\ref{Eq_13} when $\phi_{u_i}$ and $\phi_{\varepsilon}$ are determined directly from the data (Eq.~\ref{Eq_6} and \ref{Eq_12}), is able to capture well the behaviour of velocity spectra, including  the low frequency spectral variation and the shift of the spectral peak with turbulence anisotropy (Fig.~\ref{Fig5}a, \ref{Fig5}d and \ref{Fig5}g).  
For the streamwise velocity spectra, the main divergence is that the model underestimates slightly the energy peak. This is more pronounced for very anisotropic turbulence and tends to disappear as $y_B$ increases. We also note an inability of the model - constrained by its mathematical shape - to reproduce the concave region of $S_v$ at high anisotropy. 
Except for these details, we thus consider that the mathematical shape of the model is good enough to be used thereafter for all velocity components.

Because of the inability of MOST to describe properly horizontal velocity variances under unstable stratification, it was not possible to compute correct spectral behaviour for horizontal velocity spectra using the $\phi (\zeta)$ functions, when the model of \cite{olesen1984modelling} was published. Figures ~\ref{Fig5}b and \ref{Fig5}e, indeed show a huge discrepancy between the predictions of Eq.~\ref{Eq_13}  when using the classic MOST $\phi$ functions (left column in Table~\ref{table1}) and the data both in terms of low frequency spectral energy and position of the spectral peak. The discrepancy is particularly obvious for isotropic turbulence in streamwise and all anisotropy types in spanwise velocity spectra. 
Sect.~\ref{Subsection_clustering} showed that $S_u$ and $S_v$ are strongly dependant on $y_B$. Figures~\ref{Fig5}c and \ref{Fig5}f confirm that when using $\phi_{u}$, $\phi_{\varepsilon} = f(y_B, \zeta)$ (right column in Table~\ref{table1}) as parameters of the model, we are now able to better predict the shape of the spectra based only on stability $\zeta$ and turbulence anisotropy $y_B$ as input parameters. Note that due to a lack of scaling relation for $\phi_{\varepsilon_v}$ in \cite{stiperski2023generalizing} and under the commonly applied assumption of horizontal isotropy, we used $\phi_{\varepsilon_u}$ also in the spanwise velocity spectra. 
For the surface-normal spectra, the discrepancy between the data and the model based on MOST (Fig.~\ref{Fig5}h) is small, because $\phi_w = f(\zeta)$ was already a good approximation in the MOST framework (left column  in Table~\ref{table1}). When using the $\phi_w$, $\phi_{\varepsilon} = f(y_B, \zeta)$ functions (right column in Table~\ref{table1}), as parameters of the model, the spectra are, however, now even better represented (Fig.~\ref{Fig5}i). 

The model applied to temperature spectra (Eq.~\ref{Eq_14}) using Eqs.~\ref{Eq_6}, \ref{Eq_7} and \ref{Eq_16}, is shown in Fig.~\ref{Fig5}j. 
The solution based on MOST $\phi$ functions is relatively good, with mostly a slight underestimation of low frequency spectral energy for the most anisotropic bins (Fig.~\ref{Fig5}k). This is not a surprise provided that both $\phi_{\theta}$, $\phi_{\varepsilon}$ and $\phi_N$ are already well approximated by MOST. This confirms that the spread observed in Fig.~\ref{Fig3}d is mostly linked to $\zeta$. 
Furthermore, \cite{stiperski2023generalizing} showed that $\phi_{\theta}$ does not depend strongly on $y_B$ away from near-neutral stratification, while $\phi_{\varepsilon}$ has a limited dependence on $y_B$ in the stability range considered. 
If we now use $\phi_{\theta}$ and $\phi_{\varepsilon} = f(y_B, \zeta)$ and keep $\phi_N = f(\zeta)$ (cf. Table~\ref{table1}), 
the model improves for low $y_B$, while MOST scaling relations performs best for $y_B>0.45$. The reason for using MOST function for $\phi_N$ stems from no scaling relation for $\phi_N$ in \cite{stiperski2023generalizing}, as well as no clear dependence on $y_B$ is observed in the data (not shown). 

\subsection{Spectral scaling in mountainous terrain}\label{Section_complex_terrain}

\begin{figure*}
\centering
	\includegraphics[width = \linewidth]{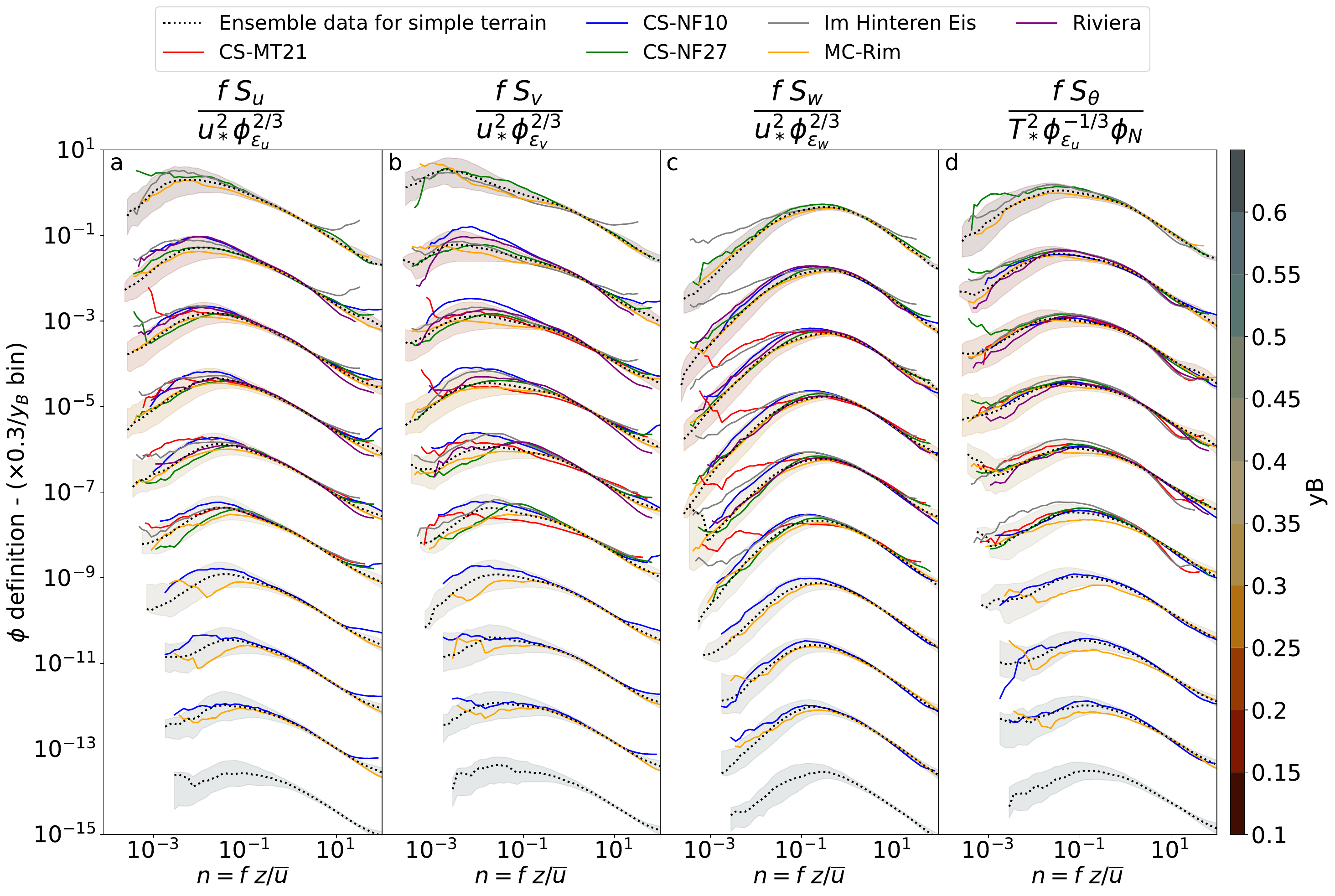}
	\caption{\label{Fig6} Scaled (a) streamwise, (b) spanwise, (c) surface-normal velocity spectra and (d) temperature spectra. In each subplot, the black dotted lines represent the median of a range of turbulence anisotropy from the ensemble data of simple terrain dataset. The shading is the interquartile range for each bin of anisotropy bin (scale on the right for the corresponding values of $y_B$). The colored solid lines are the median of the spectra per bin of $y_B$ for each of the 6 datasets included in the complex mountainous category. To facilitate the reading, spectra in each subplot are shifted down compared to each other by $0.3$, where the upper-most spectrum is in the correct position.}
\end{figure*}

For near-neutral stability conditions, \cite{tieleman1995universality} reported that the large scale behaviour of $S_u$ and $S_v$ is highly related to the upwind topography, causing differences in the spectra from one site to another, both in shape and in spectral energy. 
Over mountainous terrain, this behaviour is expected to be even more pronounced since the upwind topography is highly direction dependent. 
The scaling results (Fig.~\ref{Fig6}) show that terrain indeed influences the spectral distribution since the datasets show a large divergence amongst themselves, but also that the effect of terrain is anisotropy dependent. For the majority of the anisotropy range ($y_B>0.35$), the median spectra of six mountainous datasets mostly fall within the spread of the simple terrain datasets, despite the strong terrain complexity and the expected effect of heterogeneity on the low frequency motions. For very anisotropic turbulence ($y_B<0.3$), however, spectra of all three velocity components over complex terrain are found to always be above the simple terrain median, the spectral peak is shifted towards larger scales and the low frequency energy increases for $S_u$, $S_v$ and $S_{\theta}$ as $y_B$ decreases. The surface-normal spectra $S_w$, on the other hand, show an increase of the spectral energy around the frequency of the energy peak when approaching isotropy, as well as a secondary peak for some datasets (cf. CS-MT21). It thus appears that terrain complexity mostly impacts the surface-parallel components more for anisotropic turbulence, and surface-normal component more for isotropic turbulence.  

\begin{figure*}
\centering
	\includegraphics[width = \linewidth]{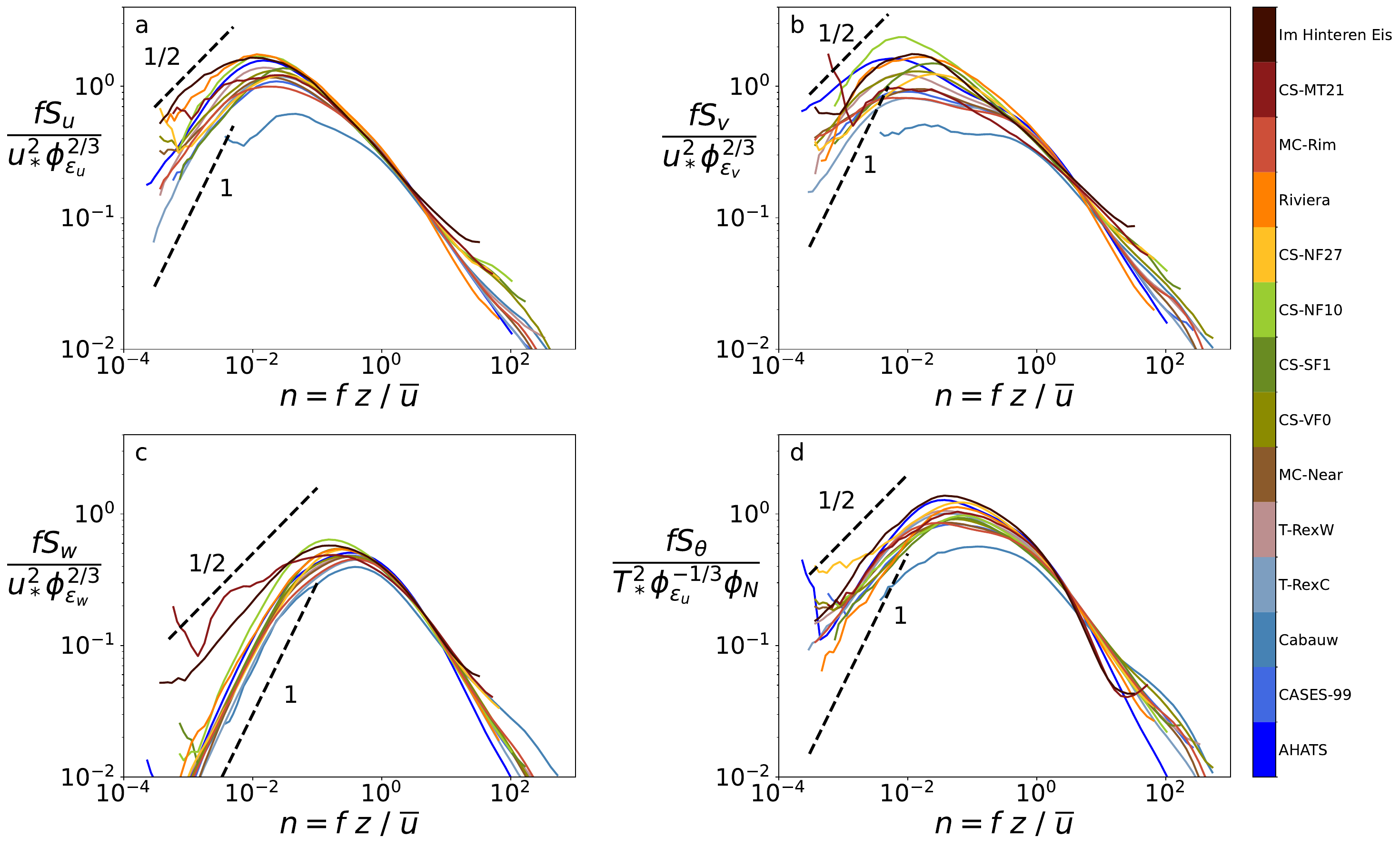}
	\caption{\label{Fig7} Median scaled spectra of (a) streamwise, (b) spanwise and (c) surface-normal velocity and (d) temperature for each individual dataset. Datasets are classified on the scale from the simplest (blue) to the most complex (red). The black dashed lines indicate spectral slopes of $1/2$ and $1$.}
\end{figure*}

According to \cite{tieleman1995universality}, the scaled complex terrain spectra are characterized by a low frequency energy higher than simple terrain spectra and a slope $\gamma$ lower than $1$ at these frequencies \citep[although what they consider as complex terrain is flat terrain with a ridge located $400$~m from the measurement site, similar to T-REX which we characterize as simple terrain in our study, ][]{perry1978horizontal, panofsky1982spectra}. 
Among the 14 datasets used in the present study, we also observe a reduction of the low frequency slope of $S_u$ and $S_v$, when spectra are plotted as the median over the whole dataset (Fig.~\ref{Fig7}). This slope decreases from nearly $1$ for some of the simple terrain down to about $0.5$ for the most complex terrain. Median temperature spectra $S_{\theta}$ show a low frequency slope lower than $1$ for all of the datasets acquired over simple and complex terrain. However, when binned according to $y_B$, low frequency slopes of both velocity and temperature spectra are closer to $1$, hence the use of $\gamma = 1$ in the analytical model. Thus the slope decrease mentioned by \cite{tieleman1995universality} and observed in Fig.~\ref{Fig7} could be the result of the averaging of spectra for data acquired over different conditions (stability, turbulence anisotropy, wind direction etc.). Interestingly, low frequency slopes of $S_w$, whether binned according to turbulence anisotropy conditions or not, are about $1$, except for the two most complex terrain datasets (Im Hinteren Eis and CS-MT21), for which $\gamma$ is lower. This feature is observed for all anisotropy conditions (Fig.~\ref{Fig6}).

\section{Discussion}\label{Section_discussion}
\subsection{Performance of the new spectral scaling}\label{Skillscore}

\begin{figure*}
\centering
\includegraphics[width = \linewidth]{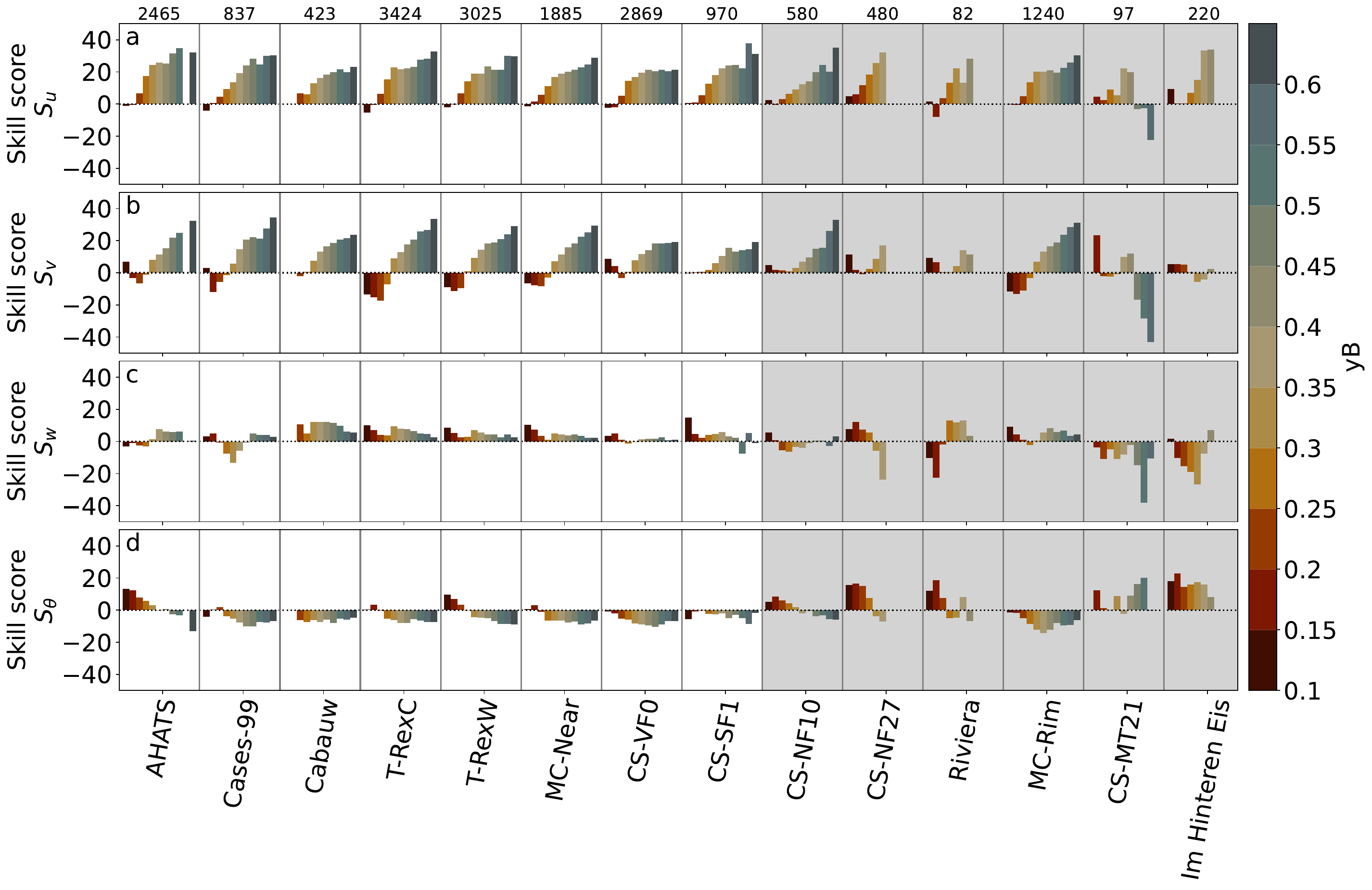}
\caption{\label{Fig8} Skill score of the improvement of the spectral model based on $\phi=f(y_B, \zeta)$ over the model using the original MOST $\phi$ functions, for (a) the streamwise velocity spectra, (b) the spanwise velocity spectra, (c) the surface-normal velocity spectra and (d) the temperature spectra for each of the dataset used in the present study. The bins color corresponds to turbulence anisotropy state. White and gray backgrounds are used for simple and complex terrain respectively. The number of $30$-min segments of each dataset is indicated at the top of each column.}
\end{figure*}

Figure \ref{Fig8} quantifies the improvement of the model brought by using the scaling relations of \cite{stiperski2023generalizing}  over using MOST (Table~\ref{table1}) for each bin of $y_B$ in each of the datasets, through the skill score $S_s$. The skill score is defined based on the mean absolute deviation ($MAD$) of the spectra, as:
\begin{equation}
    S_s = 1-\dfrac{MAD_{f(y_B, \zeta)}}{MAD_{f(\zeta)}}, 
\end{equation}

with:
\begin{equation}
    MAD = \frac{1}{T} \sum_{t=1}^{T}
    \left(\frac{1}{N} \sum_{i=1}^{N} |\log (S_{\textnormal{model}}^t(n_i)) - \log (S_{\textnormal{data}}^t(n_i))|\right),
\end{equation}
where $S_{\textnormal{model}}^t(n_i)$ and $S_{\textnormal{data}}^t(n_i)$ are the spectra from the analytical model -- either $f(y_B, \zeta)$ or $f(\zeta)$ -- and from the data respectively, at the normalized frequency $n_i$ ($N$ is the number of frequency points within each spectra) and for the time segment $t$ ($T$ is the number of time segments within each dataset). The skill score is computed for each individual spectrum and the result is the average value per bin of $y_B$.

The skill score indicates a net improvement of the ability of the model to reproduce the surface-parallel velocity spectra, by up to about $30\%$, when using $f(y_B, \zeta)$ compared to $f(\zeta)$, for both the simple and the complex terrain (Fig.~\ref{Fig8}). The negative values for very anisotropic conditions and the low values for intermediate anisotropic conditions in Fig.~\ref{Fig8} are the result of the shape constraint of the model, which does not allow to correctly capture the energy peak for the most anisotropic bins (Sect.~\ref{Subsection_modeling}). Qualitatively, looking at the individual datasets, the model is indeed still better in its new version including $y_B$: the model using $f(\zeta)$ overestimates the energy peak and underestimate the lowest frequencies, while the model using $f(y_B, \zeta)$ only overestimates the energy peak but can reproduce well the low frequencies. 
The surface-normal velocity spectra, that were found not to be strongly influenced by anisotropy, are also improved, but to a lesser degree (up to about $10\%$), for most of the simple terrain datasets. There is no coherent pattern, however, over complex terrain. 

Section~\ref{Subsection_modeling} demonstrated that the prediction of temperature spectra for the most anisotropic conditions is improved by the use of $y_B$ in the model whereas it is worsened for the less anisotropic spectra for simple terrain. Looking at each dataset separately, this behaviour is still globally valid for simple terrain but the behaviour of the most anisotropic bins is not always improved.
This is also observed for the datasets acquired over sloping terrain (CS-NF10, CS-NF27, Riviera) with a clear improvement of the prediction of very anisotropic temperature spectra by up to about $15\%$. Mountain top stations (CS-MT21, Im Hinteren Eis) show a relevant improvement for all anisotropic conditions, as well.

Interestingly, spectra computed with data from MC-Rim station (Table~\ref{table2}), located at the rim of a meteor crater, are not strongly affected by the complexity of the surrounding terrain neither for velocity components nor for temperature. This might be due to the fetch since only $25\%$ of the $30$-min segments retained after quality control (Sect.~\ref{Section_processing}) originates from the direction of the crater while the rest originate from a more uniform terrain, like MC-NEAR station.

\subsection{Discussion on the mixing layer scaling of velocity spectra}

The mixed layer scaling approach of \cite{panofsky1977characteristics} for horizontal velocity variances and its expansion to velocity and temperature spectra \citep{mcnaughton2007scaling} was developed in response to the failure of MOST. ML scaling states that the dynamics of horizontal velocity fluctuations is governed by the ML height $z_i$ as a relevant length scale. \cite{stiperski2023generalizing} showed that mixed layer scaling does collapse the surface-parallel velocity variances better than MOST, however, with a lingering dependence on anisotropy. 
We therefore explore the ML spectral scaling approach of \cite{mcnaughton2007scaling} in light of anisotropy. For this we use the data from the West tower during T-REX campaign (cf. Table~\ref{table2}), as lidar observations at this site allowed retrieving the information on $z_i$. 

\begin{figure*}
\centering
\includegraphics[width = \linewidth]{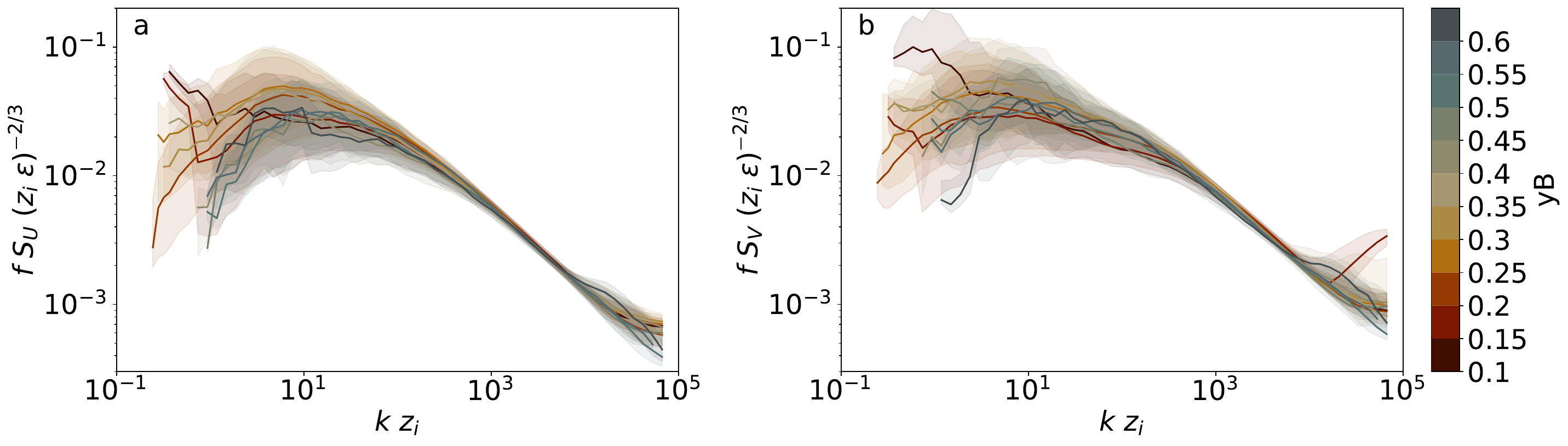}
\caption{\label{Fig9} Scaled spectra of (a) streamwise and (b) spanwise velocity component for the T-RexW dataset, scaled according to \cite{mcnaughton2007scaling} with the ML height $z_i$.}
\end{figure*}

Figure~\ref{Fig9} shows scaled streamwise and spanwise velocity spectra for which $z_i$, instead of the measurement height $z$, is taken to be the representative length scale. The clear separation of the low frequency spectral energy and the shift of the spectral peak position for surface-parallel velocity spectra with $y_B$ observed in Fig.~\ref{Fig3} nearly disappear (Fig.~\ref{Fig9}). Only a slight shift of the spectral peak position for $S_u$ is observed, by about one decade whereas it was of about three decades in Fig.~\ref{Fig3}. These results confirm the importance of ML height in the surface-parallel velocity dynamics. 

\begin{figure*}
\centering
\includegraphics[width = \linewidth]{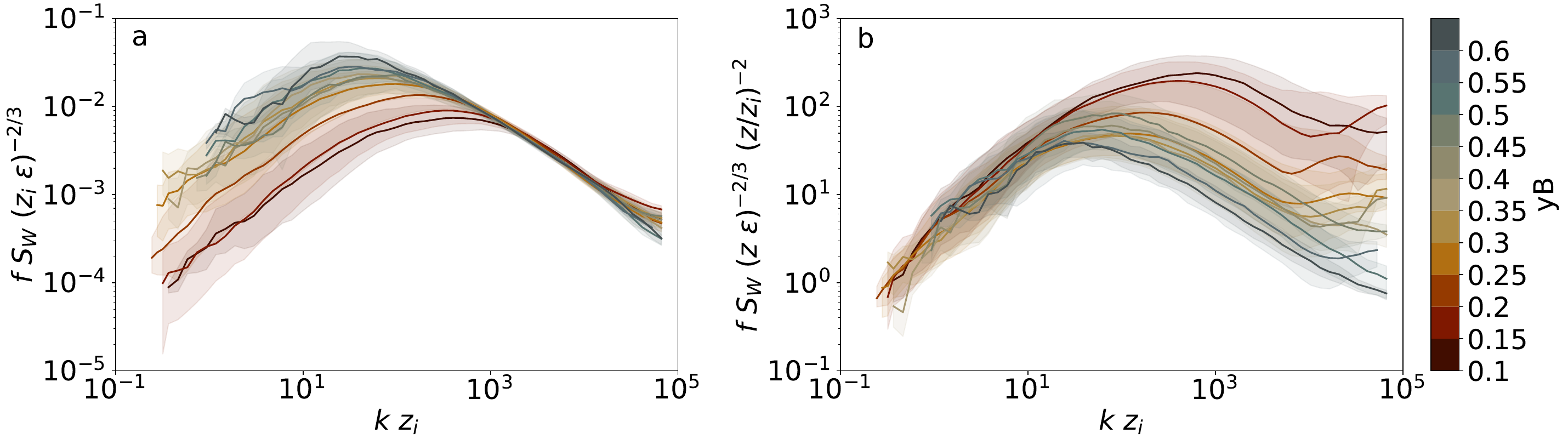}
\caption{\label{Fig11} Surface-normal velocity spectra from the T-RexW dataset, scaled according to \cite{mcnaughton2007scaling} (a) with the ML height $z_i$, and (b) using a combination of the the ML height $z_i$ and the measurement height $z$ as the representative length scales.}
\end{figure*}

As expected since $\phi_w$ and $S_w$ follow MOST, the mixed layer scaling applied to $S_w$ shows the opposite behaviour (Fig.~\ref{Fig11}a), namely an evident dependence on $y_B$ at low frequency, both in terms of spectral energy and peak position: low frequency spectral energy increases and spectral peaks are shifted towards larger scales as anisotropy decreases. This shift of the low frequency spectral energy in $S_w$ is linked to the relative height of the measurements within the mixed layer, and can be corrected by the factor $(z/z_i)^{-2}$ (Fig.~\ref{Fig11}b), derived from Fig.~5 of \cite{mcnaughton2007scaling}. However, this causes the spectra not to scale in the inertial subrange. 

The comparison between Figs.~\ref{Fig3} and \ref{Fig9} indicates that ML height does impose an important scale on surface-parallel velocity spectra, but does not integrate all of the processes affecting them.
In fact, $S_u$ and $S_v$ scaled according to ML approach, still exhibit significant scatter and a weak dependence on $y_B$, charting the limitations of the ML approach. 

\section{Conclusion}\label{Section_conclusion}

Turbulence anisotropy has been shown to improve flux-variance and flux-gradient scaling under Monin-Obukhov similarity theory (MOST) and allow it to be extended to conditions where the assumptions of MOST are clearly violated \citep{stiperski2023generalizing,mosso2023flux}. Here we expended these findings to spectral scaling in the unstable surface layer ($-100<\zeta<-0.01$) known not to obey MOST. We examined the combined role of turbulence anisotropy $y_B$ and stability $\zeta$ in explaining the dynamics of the largest turbulent scales, through the analysis of velocity and temperature spectra scaled using inertial subrange properties \citep{Kolmogorov1941a,Kolmogorov1941b} and MOST. Our analysis, based on an ensemble of 14 datasets covering a wide range of stability and surface conditions, demonstrated that turbulence anisotropy $y_B$ is able to explain the low frequency spread and the position of the spectral peak in both the scaled streamwise and spanwise velocity spectra, with a secondary effect of stability $\zeta$. 
On the other hand, stability was shown to be the primary factor able to relate the spread of the largest scales in temperature spectra and the peak position and intensity for both temperature and surface-normal velocity spectra, while $y_B$ was shown to have a secondary effect, most pronounced in convective conditions. The combination of $\zeta$ and $y_B$ in a semi-empirical model was therefore shown to be a powerful tool to describe the behaviour of the three velocity components and temperature spectra.

The spectral scaling results shed light on the nature of turbulent eddies under different states of anisotropy. Anisotropy was shown to govern the spectral peak position, so that anisotropic turbulence is characterized by eddies that peak at lower frequencies in the surface-parallel velocities and temperature spectra, and to carry larger fraction of variance than isotropic turbulence. On the other hand, isotropic turbulence, preferentially occurring in convective conditions, was shown to be comprised of eddies that are smaller in size both in spectra of streamwise and spanwise velocity and temperature, and to carry a smaller fraction of energy than anisotropic turbulence. In addition, isotropic turbulence was shown to be quasi-isotropic not only in energy but also in size of the energy containing eddies. Finally, the results highlighted that the spectral characteristics of anisotropic eddies depend on stratification, so that anisotropic turbulence under near-neutral and very unstable conditions differ. 

The complementary analysis of six datasets acquired over highly complex mountainous terrain is a first of its kind to systematically explore the impact of highly complex terrain on turbulence spectra. 
We note a net improvement of the ability to describe the surface-parallel velocity spectra $S_u$ and $S_v$ in complex terrain when incorporating $y_B$ into the analysis, compared to when using only $\zeta$. Complexity of turbulence in velocity and temperature spectra are quite well predicted through the semi-empirical model, including $y_B$ and $\zeta$ as parameters, for data acquired on steep slopes. For mountain top locations, however, the spectral model has only limited ability. The cause of this might stem from the use of double rotation, which might incorrectly assign the velocity variances compared to the gravity vector especially under conditions of flow separation. In addition, other factors not taken into account might play a role, such as for example streamline curvature.

In the end, we explored the contribution of the mixed layer scaling to the understanding of the largest scales of turbulence for velocity spectra, and found that, as previously observed, including the mixed layer height instead of measurement height into scaling reduced the low frequency spread of $S_u$ and $S_v$. This approach, however, still exhibited large scatter and a weak dependence on anisotropy, and therefore did not outperform the Kaimal scaling with included anisotropy.

The present study highlights the advantage of examining a wide range of stability and surface conditions in gaining insights into the complex surface-layer dynamics. Thus, the clear dependence of temperature spectra on stability $\zeta$, not observed by \cite{kaimal1972spectral}, is here shown to emerge out of the extensive stability range probed.
The results further highlight that turbulence anisotropy, quantified through the invariant $y_B$, is not only a powerful tool in expanding flux-variance, flux-gradient and spectral scaling to non-canonical conditions, but through the modified spectral scaling offers a deeper understanding of the nature and physical processes driving surface-layer turbulence, both over canonical and complex surfaces.

\section*{Acknowledgements}

The providers of various data sets are acknowledged: the managers of the database of the Cabauw Experimental Site for Atmospheric
Research (CESAR), UCAR/NCAR - Earth Observing Laboratory (EOL) for the access to the AHATS, CASES-99, METCRAX II and T-REX data sets, the University of Basel and ETH for the Riviera dataset, and all the collaborators of Department of Atmospheric and Cryospheric Sciences (ACINN), who worked for many years to make the reliable and continuous database of the i-Box study area, and Im Hinteren Eis station available.

\section*{Conflict of interest}
The authors declare no conflict of interest.


\graphicalabstract{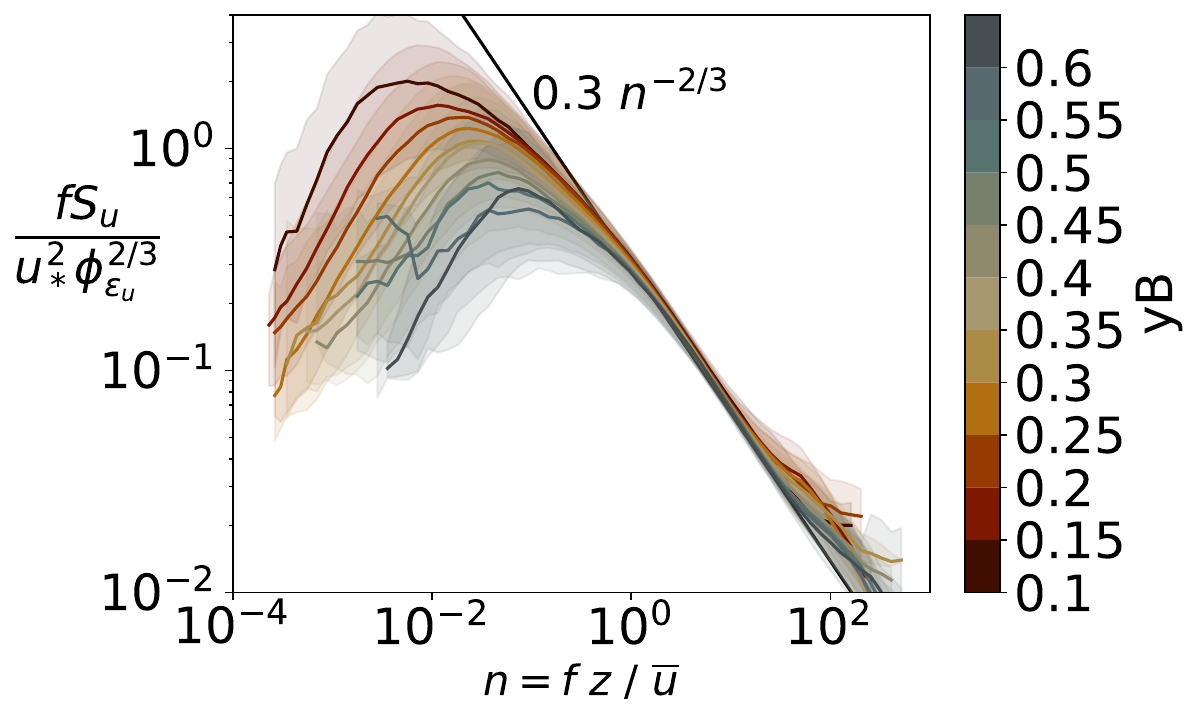}{Turbulence anisotropy has been shown to improve flux-variance and flux-gradient scaling, extending them to conditions where the assumptions of Monin-Obukhov similarity theory (MOST) are violated. Our work expends on these findings to spectral scaling in the unstable surface layer. Turbulence anisotropy, together with the stability parameter, is able to explain the low frequency spectral energy and the position of the spectral peak for velocity and temperature spectra. The novel spectral scaling elucidates on the nature of anisotropic eddies.}

\end{document}